\documentclass[11pt,A4,twoside]{article}
\usepackage[latin1]{inputenc}
\usepackage{amssymb}
\usepackage{amsmath}
\usepackage{amsfonts}
\usepackage{amsbsy}
\usepackage{natbib}
\usepackage{eucal}
\usepackage{graphicx}
\usepackage{titlesec}
\usepackage[english]{babel}
\usepackage{times}
\usepackage{titling}
\usepackage{mathptmx}
\usepackage{float}
\usepackage[font=small]{caption}
\usepackage{array}
\usepackage[colorlinks,citecolor=blue,urlcolor=blue]{hyperref}
\usepackage{authblk}
\usepackage{amsthm}

\usepackage{setspace}
\DeclareSymbolFont{txgreek}{OML}{cmr}{m}{it}

\renewcommand{\abstract}[1]{{\small\noindent
\hrulefill\par \vspace*{0.1cm}\noindent{\small\bf\sffamily
{Abstract}}\parindent=0pt\par\noindent\vspace{-0.1cm}\noindent\hrulefill\par\vspace*{0.5\baselineskip}\hspace*{0cm}\renewcommand{\baselinestretch}{1.1}\sffamily{#1}\par\vspace*{-0.1cm}\noindent\hrulefill}}

\def\and{,\;}
\DeclareMathSizes{11}{11}{7.8}{6.5}

\def\paragraf{\fontsize{9}{10pt}\fontfamily{phv}\fontshape{it}\selectfont}
\titleformat{\paragraph}
{\titlerule[0pt]
\vspace{0cm}%
\paragraf}
{\theparagraph}{.5em}{\vspace*{0\baselineskip}}

\def\titol{\fontsize{12.045}{12pt}\fontfamily{phv}\fontseries{b}\selectfont}
\titleformat{\section}
{\titlerule[0pt]
\vspace{0.2cm}%
\titol}
{\thesection.}{.5em}{\vspace*{0\baselineskip}}

\def\titolp{\fontsize{11.045}{11pt}\fontfamily{phv}\fontseries{b}\fontshape{it}\selectfont}
\titleformat{\subsection}
{\titlerule[0pt]\bigskip
\vspace{-0.4cm}%
\titolp}
{\thesubsection.}{.5em}{\vspace*{0\baselineskip}}

\def\titolpp{\fontsize{10.045}{10pt}\fontfamily{phv}\fontshape{it}\selectfont}
\titleformat{\subsubsection}
{\titlerule[0pt]
\vspace{-0.4cm}%
\titolpp}
{\thesubsubsection.}{.5em}{\vspace*{0\baselineskip}}

\usepackage{titling}
    \pretitle{\begin{center}\sffamily\fontsize{18pt}{20pt}\selectfont}
    \posttitle{\par\end{center}}
    \preauthor{\begin{center}\fontsize{12pt}{14pt}\selectfont}
    \postauthor{\par\end{center}\vspace{0bp}}
    \predate{}
    \date{}
    \postdate{}

\title{Compositional Periodic Spline Approximation for Circular Density Data in Bayes Spaces}

\thanksmarkseries{arabic}
\author{Jitka Machalov{\'a}\thanks{\parbox[t]{0.95\textwidth}{\raggedright\footnotesize\setlength{\baselineskip}{8pt} Department of Mathematical Analysis and Applications of  Mathematics, Faculty of Science, Palack{\'y} University, Olomouc, Czech Republic}} \and  Jana Heckenbergerov{\'a}\thanks{\parbox[t]{0.95\textwidth}{\raggedright\footnotesize\setlength{\baselineskip}{8pt} Institute of Mathematics and Quantitative Methods, Faculty of Economics and Administration,   University of Pardubice, Czech Republic}} \and Karel Hron\thanks{\parbox[t]{0.95\textwidth}{\raggedright\footnotesize\setlength{\baselineskip}{8pt}Department of Mathematical Analysis and Applications of  Mathematics, Faculty of Science, Palack{\'y} University, Olomouc, Czech Republic}}}

\def\headers#1{\fontsize{8.5}{10}\centering\sffamily\itshape{#1}}
\def\page#1{\fontsize{8.5}{10}\sffamily{#1}}
\usepackage{fancyhdr}%Headers

\begin{document}
\maketitle

% Headers
\thispagestyle{empty}
\renewcommand{\headrulewidth}{0truecm}
\pagestyle{fancy}
\rhead[\headers{Compositional Periodic Spline Approximation for Circular Density Data in Bayes Spaces}]{\page{\thepage}}
\lhead[\page{\thepage}]{\headers{Jitka Machalov\'{a} \and  Jana Heckenbergerov\'{a} \and Karel Hron}}
 \lfoot{} \rfoot{}
\cfoot{}

\abstract{This paper proposes a novel framework for the approximation and analysis of circular density data using compositional periodic splines within Bayes spaces with the Hilbert space structure. By applying the centered log-ratio transformation, densities are represented in a subspace of the standard $L^2$ space of real-valued functions, which enables the use of functional data analysis tools while preserving the relative nature of distributions and their periodic structure. A coefficient-based construction of periodic splines with a zero-integral constraint is developed, together with matrix formulations for both smoothing splines and penalized splines, allowing efficient estimation and implementation. The methodology is applied to long-term wind direction data, where it provides smooth and interpretable density estimates and supports further statistical analysis, including functional regression. The results demonstrate the practical relevance of the proposed approach and its potential for extensions to more complex density-valued data.}

%\paragraph{MSC: \textcolor{red}{NUTNO DOPLNIT}} 

\paragraph{Keywords: circular data; density estimation; Bayes spaces; periodic splines; functional data analysis; wind direction}

\renewcommand{\baselinestretch}{1.2}
\bigskip

\section{Introduction}

\label{sec:1}

Directional distribution modeling is essential for the analysis of data with inherent periodicity, such as wind direction, animal movement, and time-of-day patterns. The statistical techniques employed in this context address the specific challenges posed by the circular nature of such data, where traditional models fail to capture the underlying structure. More broadly, periodic data modeling plays a crucial role in the analysis of datasets exhibiting regular and repeating temporal patterns. These types of data are widely encountered across diverse fields, including meteorology, economics, and biomedical sciences. 

With the increasing availability of large-scale circular and directional datasets, the development of methods for approximation and statistical modeling their underlying distributions has become increasingly important. A fundamental step in the analysis of distributions of circular data is their appropriate approximation. Among the available approaches, restricted cubic splines have gained considerable attention due to their ability to flexibly model nonlinear relationships while incorporating constraints that reflect periodicity. In \cite{lusa20}, this approach was shown to outperform alternative methods such as the cosinor model and standard cubic splines. In particular, restricted cubic splines effectively capture seasonal variations and hormonal fluctuations while reducing variability at the boundaries of the period.

The general objective of directional data analysis is to approximate the underlying directional distribution and subsequently investigate its specific properties. This task has important practical implications. For example, \cite{lin19} developed a trivariate statistical model for assessing wave energy resources based on significant wave height, mean wave period, and direction. Similarly, \cite{zerbe22} applied optimized hot spot and directional distribution analyses to characterize the spatio-temporal variability of large wildfires, demonstrating the broad applicability of such methods in environmental studies.

Recently, a number of advanced directional statistical models have been proposed, including kernel-based approaches and parametric copula models. For instance, \cite{benlakhdar22} introduced a hierarchical von Mises distribution model, providing a flexible framework for modeling complex directional datasets with multiple modes and asymmetries. The work of \cite{DiMarzio22} addresses density estimation for circular data subject to measurement errors by proposing kernel-based techniques that improve estimation accuracy. Furthermore, \cite{carta20} developed a global sensitivity analysis framework for wind farm power output estimation using Shapley effects and vine copulas to capture dependencies among input variables. These methods enable the estimation of long-term wind characteristics at a target site, which is essential for optimal wind turbine placement. In \cite{wang21}, a modeling framework for three-dimensional circular-linear-linear (C-L-L) data is proposed, combining parametric copula-based models with nonparametric kernel density estimation. Collectively, these contributions highlight the importance of incorporating directionality into statistical modeling, leading to improved understanding and prediction of phenomena with inherent circular characteristics.

The analysis of wind direction represents a particularly important application in meteorology and engineering. The study by \cite{weber97} emphasizes the role of the standard deviation of horizontal wind direction in atmospheric turbulence and dispersion modeling, showing that estimators based on sine and cosine transformations are the most efficient. In \cite{droppo08}, directional bias in wind rose generation and sector-based air dispersion modeling is addressed through a revised data-transfer algorithm that ensures more accurate representation of wind direction data. Additionally, \cite{kamisan10} compare several circular probability distributions for modeling southwest monsoon wind direction in Malaysia, employing mean circular distance and chord length as performance criteria.

Beyond wind direction alone, multiple environmental variables can be analyzed jointly within a directional framework. The work of \cite{Ye19} extends the REBMIX algorithm to the von Mises family in order to model the joint distribution of wind speed and direction, successfully capturing multimodal characteristics. The joint analysis of wind and wave conditions is particularly important for offshore structural design, where advanced statistical models have been developed to account for complex dependencies and directional effects \citep{ditlevsen02, horn18, lin19, vanem23, vanem24, kaliske24}. Moreover, the study of extreme wind properties is crucial for structural engineering and risk assessment. Existing approaches focus on directional dependence, robust estimation techniques, and the influence of wind direction on structural response \citep{coles94, palutikof99, MacDonald11, holmes20, cui24}. Together, these studies contribute to a deeper understanding of joint wind characteristics and support improved structural design and safety evaluation.

From another perspective, samples of directional distributions can be interpreted as functional objects in Bayes spaces. Initial developments in this direction are presented in \cite{machalova2025}. Within this framework, the approximation of distributions represents a fundamental step, enabling their subsequent analysis using tools from functional data analysis. The spline-based approximation naturally emerges as a suitable approach; however, it must be carefully adapted to respect the intrinsic properties of directional distributions, in particular their relative nature and periodicity constraints.

Building on this perspective, the main objective of this paper is to develop a novel methodology for the approximation and analysis of circular data based on compositional periodic splines in Bayes spaces. The proposed approach integrates periodic spline smoothing with the geometry of Bayes spaces, allowing for a consistent treatment of density functions as relative objects while simultaneously preserving their periodic structure. In particular, we introduce a coefficient-based construction of periodic smoothing splines with a zero-integral constraint in a subspace of the standard $L^2$ space of real functions, which enables efficient computation and direct applicability of functional data analysis techniques. 

The remainder of the paper is organized as follows. Section \ref{sec:2} introduces the fundamentals of functional data analysis in Bayes spaces, including the representation of density functions and the centered log-ratio transformation. Section \ref{sec:3} provides an overview of periodic (circular) distributions and their key properties. In Section \ref{sec:4}, we develop the proposed methodology of compositional periodic spline approximation, including the construction of periodic smoothing splines and their penalized variants. Section \ref{sec:5} presents an application to real wind direction data, demonstrating the practical performance of the proposed approach and its use within a functional data analysis framework. Finally, Section \ref{sec:6} concludes the paper and outlines the directions for future research.

\section{Density Data Analysis in Bayes Spaces}
\label{sec:2}

The distributional properties of large-scale database systems can be characterized by a cumulative distribution function or by a probability density function. Densities on a given domain are positive functions integrating to one, which limits their direct analysis using standard functional data analysis (FDA) methods that assume unconstrained real-valued functions \citep{ramsay05}. This constraint also affects spline-based approximations commonly used in FDA. More importantly, densities exhibit scale invariance and relative scale properties, distinguishing them from standard $L^2$ functions \citep{egozcue06, boogaart10, boogaart14}. 

For bounded domains $I = [a, b] \subset \mathbb{R}$, the Bayes space $\mathcal{B}^2(I)$ models positive densities with their square-integrable logarithm. This space, with a separable Hilbert structure, treats densities as scale-invariant, meaning that any positive multiple of a density holds the same relative information \citep{boogaart14}. This differs from $L^2(I)$, as $\mathcal{B}^2(I)$ consists of equivalence classes of proportional densities. The vector space structure in $\mathcal{B}^2(I)$ is defined by \textit{perturbation} and \textit{powering} for densities $f, g \in \mathcal{B}^2(I)$ and $\alpha \in \mathbb{R}$
\begin{equation*}
(f \oplus g)(x) = \frac{f(x)g(x)}{\int_I f(y)g(y)\,\mathrm{d}y}, \quad (\alpha \odot f)(x) = \frac{f(x)^\alpha}{\int_I f(y)^\alpha\,\mathrm{d}y}, \quad x \in I,
\end{equation*}
where, due to their scale invariance, rescaling to the unit integral constraint is just optional. The perturbation-difference between two densities is defined as $f \ominus g = f \oplus [(-1) \odot g]$ and the inner product completing the Hilbert space as
\begin{equation*}
\langle f, g \rangle_{\mathcal{B}} = \frac{1}{2\eta} \int_I \int_I \ln \frac{f(x)}{f(y)} \ln \frac{g(x)}{g(y)} \, \mathrm{d}x \, \mathrm{d}y,
\end{equation*}
with $\eta = b - a$. The corresponding norm and distance are
\begin{equation*}
\|f\|_{\mathcal{B}} = \sqrt{\langle f, f \rangle_{\mathcal{B}}}, \quad d_{\mathcal{B}}(f, g) = \|f \ominus g\|_{\mathcal{B}}.
\end{equation*}
An isometric isomorphism between $\mathcal{B}^2(I)$ and $L^2(I)$ is achieved using the centered log-ratio (clr) transformation \citep{boogaart14}
\begin{equation*}
\mathrm{clr}(f)(x) = \ln f(x) - \frac{1}{\eta} \int_I \ln f(y) \, \mathrm{d}y\qquad \forall x\in I.
\end{equation*}
This transformation allows for standard operations
\[
\mathrm{clr}(f \oplus g)(x) = \mathrm{clr}(f)(x) + \mathrm{clr}(g)(x), \quad \mathrm{clr}(\alpha \odot f)(x) = \alpha \cdot \mathrm{clr}(f)(x),
\]
and the inner product becomes
\[
\langle f, g \rangle_{\mathcal{B}} = \langle \mathrm{clr}(f), \mathrm{clr}(g) \rangle_{L^2} = \int_I \mathrm{clr}(f)(x)\, \mathrm{clr}(g)(x) \, \mathrm{d}x.
\]
The clr transformation imposes the constraint
\begin{equation*}
\int_I \mathrm{clr}(f)(x) \, \mathrm{d}x = 0,
\end{equation*}
defining the clr space $L_0^2(I)$, a subspace of $L^2(I)$. The inverse clr transformation is given by
\[
\mathrm{clr}^{-1}(z)(x) = \frac{\exp(z(x))}{\int_I \exp(z(y))\,\mathrm{d}y}, \qquad z \in L_0^2(I), \; x \in I,
\]
which ensures that the resulting function is a valid density in $\mathcal{B}^2(I)$.

\subsection{Descriptive statistics of density data}

Let $y_1, \dots, y_n \in \mathcal{B}^2(I)$ denote a sample of density functions, and let $z_i(x) = \mathrm{clr}(y_i)(x) \in L_0^2(I)$ be their clr representations. Since $L_0^2(I)$ is a Hilbert space, standard descriptive statistics can be defined in this transformed space.
The sample mean function is given by
\[
\bar{z}(x) = \frac{1}{n} \sum_{i=1}^n z_i(x), \quad x \in I,
\]
which satisfies the zero-integral constraint and thus belongs to $L_0^2(I)$. The corresponding mean density in $\mathcal{B}^2(I)$ is obtained via the inverse clr transformation as
\[
\bar{y}(x) = \mathrm{clr}^{-1}(\bar{z})(x).
\]
and corresponds to the mean function formulated directly in the Bayes space \citep{boogaart14},
\[
\bar{y}(x) = \frac{1}{n} \odot\bigoplus_{i=1}^n y_i(x).
\]
The variability of the sample can be described by the functional variance
\[
\mathrm{Var}(z)(x) = \frac{1}{n} \sum_{i=1}^n \left(z_i(x) - \bar{z}(x)\right)^2,
\]
and the associated functional standard deviation
\[
\mathrm{sd}(z)(x) = \sqrt{\mathrm{Var}(z)(x)}.
\]
These characteristics provide pointwise measures of variability in the clr space and are commonly used for graphical and numerical summaries of functional data.

\subsection{Function-on-scalar regression model for density responses}
\label{subsec:funcreg}

A number of contributions have been published about functional regression models in $L^2(I)$. We will follow the key concepts presented in \cite{talska18}, where function-on-scalar linear regression with the density response variable is addressed.

Let $y_i \in \mathcal{B}^2(I)$, $i=1,\dots,n$, denote density responses, and let $x_{ij}$, $j=1,\dots,p$, be scalar covariates. The function-on-scalar regression model in Bayes spaces is defined as
\[
y_i = \beta_0 \oplus \bigoplus_{j=1}^p (x_{ij} \odot \beta_j) \oplus \varepsilon_i,
\]
where $\beta_0,\, \beta_j \in \mathcal{B}^2(I)$ are functional regression parameters and $\varepsilon_i$ are random errors in $\mathcal{B}^2(I)$ with zero mean. Applying the clr transformation yields an equivalent linear model in $L_0^2(I)$
\[
\mathrm{clr}(y_i)(x) = \mathrm{clr}(\beta_0)(x) + \sum_{j=1}^p x_{ij}\,\mathrm{clr}(\beta_j)(x) + \mathrm{clr}(\varepsilon_i)(x),
\]
which allows for the use of standard functional regression techniques.

In practice, both the response functions and regression coefficients are represented using a suitable basis expansion, typically $B$-splines, so that the problem reduces to the estimation of a multivariate linear model for the corresponding basis coefficients. The estimation is commonly performed by least squares minimization in the $L^2$ space, possibly combined with smoothing penalties to control the roughness of the estimated functions.
\section{Periodic Distributions and Their Properties}
\label{sec:3}

The field of directional statistics has received considerable attention over the past three decades. An overview of its development can be found in \cite{mardia75} and \cite{fisher93}, followed by the influential monographs \cite{mardia00} and \cite{jamma01}. The introduction of the first \texttt{R} package for circular data analysis is presented in \cite{pewsey13}. More recent theoretical developments are summarized in \cite{ley17}, while the companion book \cite{ley19} focuses on modern datasets and their analysis.

Periodic data arise in many scientific contexts, both from angular measurements and from daily or seasonal activity patterns. Examples include bond angles measured in molecules, longitudes of earthquake epicenters, times of day at which emergency calls are received, or seasonal variation in global energy consumption or television viewing figures. The natural graphical representation of such data consists of points located along the circumference of a circle. Importantly, circular variables are inherently periodic, and their origin (or zero point) is defined arbitrarily rather than emerging naturally from the system. Once an origin (e.g. North) and an orientation (e.g. clockwise) are chosen, a circular observation can be represented by an angle $\theta$ measured along the unit circle. Negative values of $\theta$ correspond to measurements taken in the direction opposite to the chosen orientation \citep{pewsey13}.

To standardize the representation, angles are typically measured in radians from the positive horizontal axis in a counterclockwise direction. Thus, angles are implicitly considered modulo $2\pi$. For this choice of unit, origin, and orientation, the unit vector $\mathbf{x}$ and the angle $\theta$ are related through $\mathbf{x} = (\cos{\theta},\sin{\theta})^{\top}$. Further simplification is achieved by representing $\mathbf{x}$ in the complex plane. A circular observation can then be expressed as the complex number $z = e^{i\theta} = \cos{\theta} + i\sin{\theta}$, where $i = \sqrt{-1}$.

\begin{figure}[htbp]
    \centering
    \includegraphics[width=0.49\textwidth]{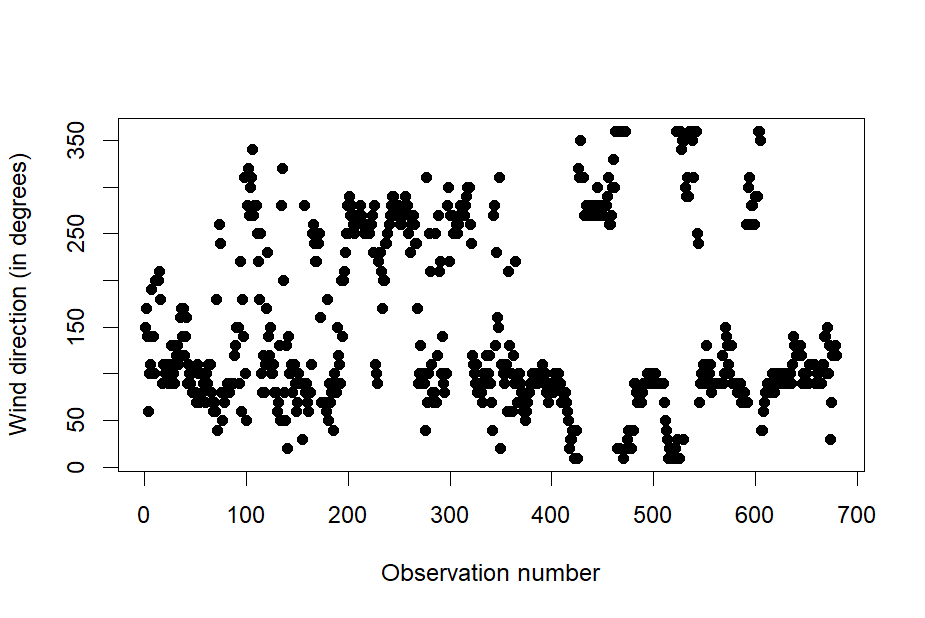}
    \includegraphics[width=0.49\textwidth]{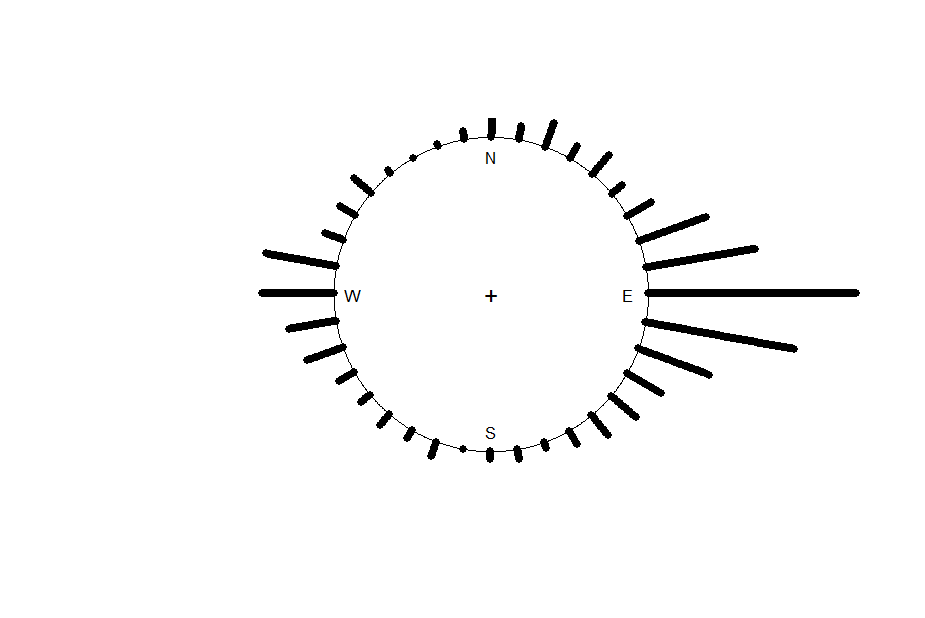}
      \caption{Linear plot (left) and raw circular data plot (right) of wind directions recorded at Pardubice Airport in January 2014}
        \label{fig:Jan14_data} 
\end{figure}

\begin{figure}[htbp]
    \centering
    \includegraphics[width=0.49\textwidth]{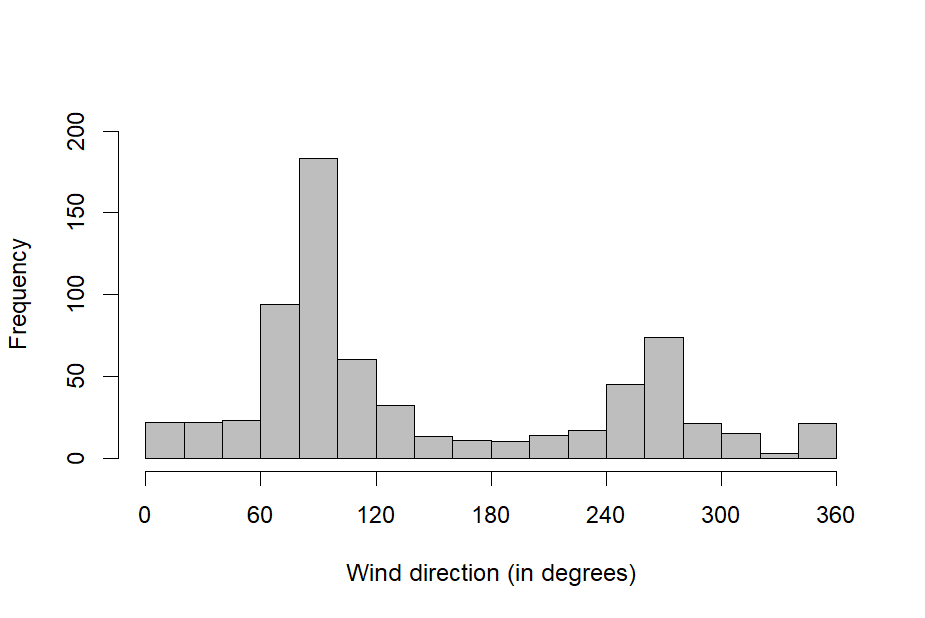}
    \includegraphics[width=0.49\textwidth]{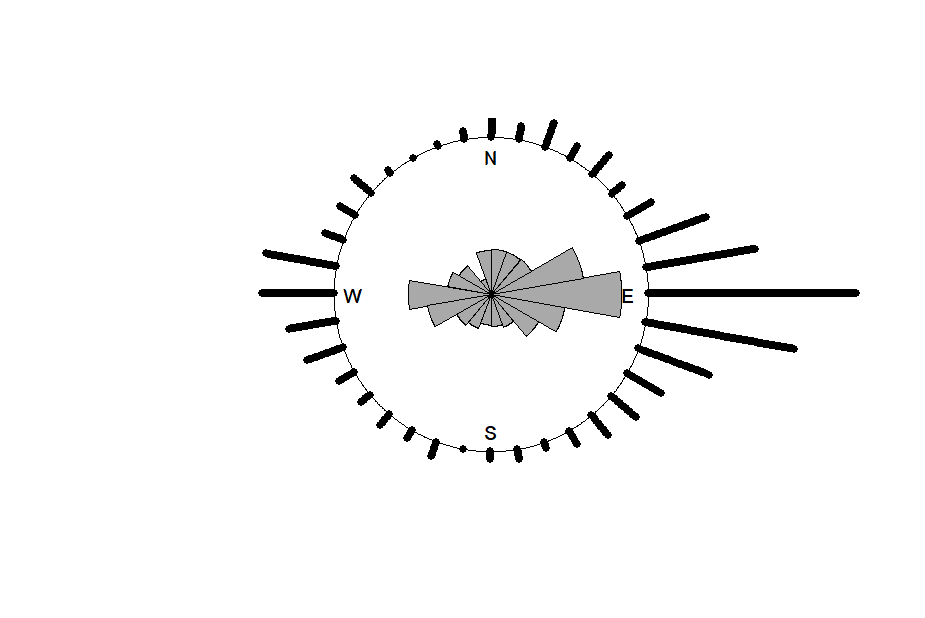}
      \caption{Linear histogram (left) with the cut-point at 0 (north) and raw circular data plot with a rose diagram (right) for the wind directions recorded at Pardubice Airport in January 2014}
        \label{fig:Jan14_hist} 
\end{figure}

Wind direction datasets typically contain angles measured in degrees clockwise from the north. Graphical representations of such circular data are shown in Figure \ref{fig:Jan14_data}. As mentioned earlier, the most natural representation is a point on the unit circle. A circular histogram can be constructed as a standard linear histogram by cutting the circle at a specified point (e.g. at 0 corresponding to North). However, this approach removes the inherent periodicity, which may hinder interpretation. Implicitly, the reader is expected to mentally wrap the histogram back onto the circle. The rose diagram is a related visualization, in which relative frequencies are represented by sector areas rather than bars. Figure \ref{fig:Jan14_hist} illustrates both a histogram and a rose diagram for the selected wind direction dataset.

\subsection{Measures of location, concentration and dispersion}

\begin{figure}[h!]%[htbp]
    \centering
    \includegraphics[width=0.99\textwidth]{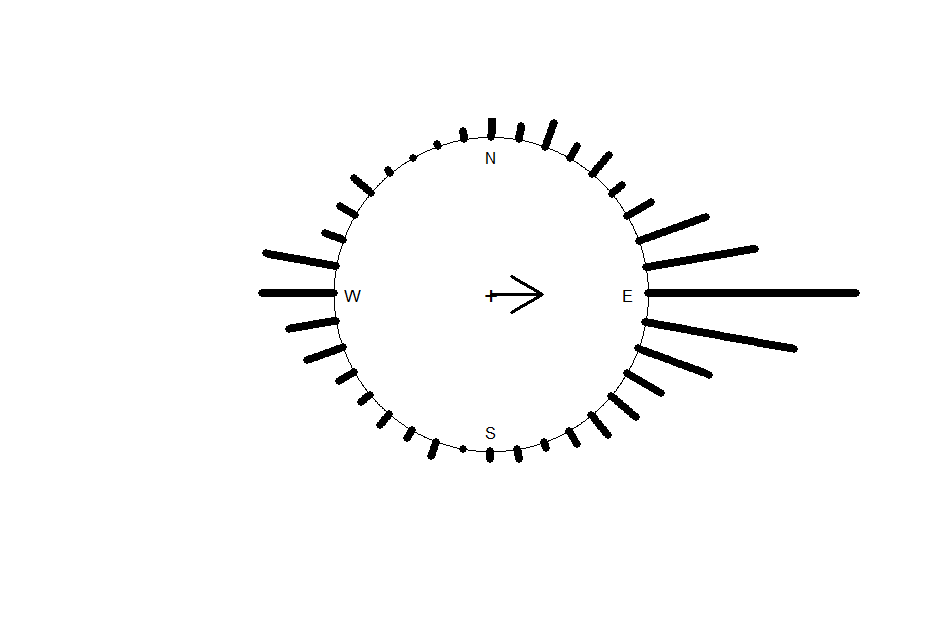}
    \caption{Raw circular data plot together with bold arrow identifying the mean resultant vector for the wind directions recorded at Pardubice Airport in January 2014}
        \label{fig:Jan14_mean} 
\end{figure}

Let us consider a random sample of circular observations of size $n$, with associated unit vectors $\mathbf{x}_1 , \dots , \mathbf{x}_n$, angles $\theta_1 , \dots ,\theta_n$, and complex numbers $z_1, \dots ,z_n$. For $p=0, \pm 1, \pm 2, \dots$, the \textit{$p$-th sample trigonometric moment about the zero direction} is defined as
\begin{equation*}
t_{p,0} = \frac{1}{n}\sum_{j=1}^{n} z_{j}^{p}
= \frac{1}{n}\sum_{j=1}^{n} e^{ip\theta_{j}}
= \frac{1}{n}\sum_{j=1}^{n} \left( \cos p\theta_{j} + i \sin p\theta_{j}  \right)
= a_{p} + i b_{p},
\end{equation*}
where
\begin{equation*}
a_{p} = \frac{1}{n}\sum_{j=1}^{n} \cos p\theta_{j},
\quad
b_{p} = \frac{1}{n}\sum_{j=1}^{n} \sin p\theta_{j}.
\end{equation*}
The complex number $t_{p,0}$ defines the \textit{$p$-th mean resultant vector} in the complex plane, characterized by the \textit{mean resultant length} $\bar{R_{p}}$ and the \textit{mean resultant direction} $\bar{\theta_{p}}$. When $\bar{R_{p}} > 0$, the polar representation is
\begin{equation*}
t_{p,0} = \bar{R_{p}}  e^{i\bar{\theta_{p}}} 
= \bar{R_{p}} \left( \cos \bar{\theta_{p}} + i \sin \bar{\theta_{p}}  \right),
\end{equation*}
with
\[
a_{p} = \bar{R_{p}} \cos \bar{\theta_{p}}, 
\quad 
b_{p} = \bar{R_{p}} \sin \bar{\theta_{p}}.
\]
For $p=1$, $t_{1,0}$ is referred to as the \textit{sample mean vector}, and $\bar{\theta_{1}}, \bar{R_{1}}$ are denoted as the \textit{sample mean direction} $\bar{\theta}$ and the \textit{sample mean resultant length} $\bar{R}$, respectively. Figure \ref{fig:Jan14_mean} illustrates the sample mean vector for the previously introduced dataset. Numerically, it corresponds to $\bar{\theta} = 90.172$, $\bar{R} = 0.3252$, $a_{1} = 0.3252$, and $b_{1} = -0.00098$. For unimodal and approximately symmetric samples, $\bar{\theta}$ provides a meaningful measure of central tendency. The quantity $\bar{R}$ is the most commonly used measure of concentration. The \textit{sample circular variance} is defined as $V = 1 - \bar{R}$, and the \textit{sample circular standard deviation} is given by
\[
\hat{\sigma} = \sqrt{-2 \log (1 - V)}.
\]
For highly concentrated samples with small $V$, the approximation $\hat{\sigma} \doteq \sqrt{2V} = \sqrt{2(1-\bar{R})}$ is often used, commonly referred to as the \textit{mean angular deviation} \citep{pewsey13}.

\subsection{Circular Distributions}

By analogy with random variables on the real line, a circular probability distribution can be specified via its distribution function. However, two important complications arise due to the periodic nature of circular variables. First, the definition of a circular distribution depends on the choice of initial direction, orientation, and measurement units. Second, the equivalence of angles $\theta$ and $\theta + 2k\pi$, $k \in \mathbb{Z}$, implies that circular distributions must be periodic. It is therefore customary to define the \textit{circular distribution function} $F$ as
\begin{equation*}
F(\theta) = P(0 < \Theta \leq \theta), \quad 0 \leq \theta \leq 2\pi,
\end{equation*}
with the periodicity condition
\begin{equation*}
F(\theta + 2\pi) - F(\theta) = 1, \quad -\infty < \theta < \infty.
\end{equation*}
If $F$ is absolutely continuous, it admits a corresponding \textit{circular probability density function} $f$ such that
\begin{equation*}
\int_{\phi}^{\psi} f(\theta) \, d\theta = F(\psi) - F(\phi), \quad -\infty < \phi \leq \psi < \infty.
\end{equation*}
A function $f$ is a valid circular density if and only if
\begin{itemize}
    \item[1)] $f(\theta) \geq 0$ almost everywhere;
    \item[2)] $f(\theta + 2\pi) = f(\theta)$ almost everywhere;
    \item[3)] $\int_{0}^{2\pi} f(\theta) \, d\theta = 1$.
\end{itemize}
The first condition is standard, while the remaining two reflect periodicity.

Various general methods exist for constructing circular distributions. One approach is \textit{perturbation}, which yields distributions such as the cardioid or sine-skewed distributions. Another approach is \textit{wrapping}, where a real-valued random variable $X$ is mapped to $\Theta = X \ (\mathrm{mod}\ 2\pi)$. The wrapped Cauchy and wrapped normal distributions are common examples. A third method is \textit{transformation of argument}, replacing $\theta$ by a function of itself in an existing density. Finally, \textit{projected} or \textit{offset} distributions arise by integrating a joint density $f(r,\theta)$ over the radial component $r$.

\begin{figure}[h!]%[htbp]
    \centering
    \includegraphics[width=0.99\textwidth]    {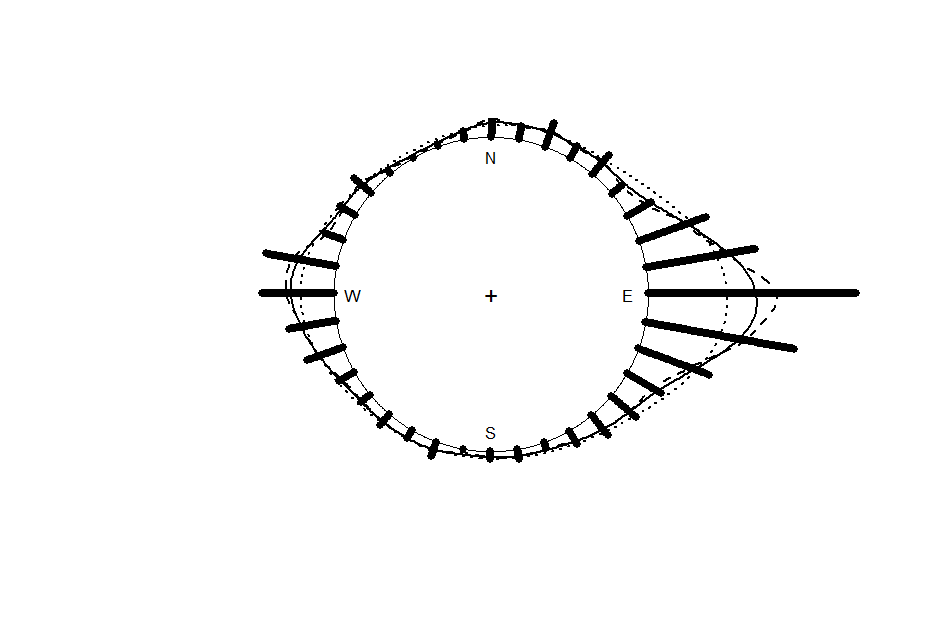}
    \caption{Raw circular data plot with three von Mises kernel density estimates with bandwidths of 10 (dotted), 40 (solid) and 120 (dashed) for the wind directions recorded at Pardubice Airport in January 2014}
        \label{fig:Jan14_kernel} 
\end{figure}

The von Mises distribution is often referred to as the \textit{circular normal} distribution due to its appealing mathematical properties. It was introduced in 1918, and \cite{mardia00} describes several constructions leading to it. The distribution is unimodal and symmetric about its mean direction, with a normalizing constant involving a modified Bessel function. In recent years, kernel density estimation has become a popular tool for graphical representation of circular data, as illustrated in Figure \ref{fig:Jan14_kernel}. A natural choice of kernel is the von Mises density, which is commonly used as the default option in circular data analysis software \citep{pewsey13}.
\section{Periodic Smoothing Splines}
\label{sec:4}

In the analysis of circular and periodic data, spline-based approximations provide a flexible and computationally convenient tool for obtaining smooth representations of underlying density functions. In our setting, however, the approximation must simultaneously respect two essential features: periodicity of the support and the zero-integral constraint induced by the clr representation in $L_0^2(I)$. This section, therefore, develops the construction of periodic smoothing splines adapted to these requirements. We first recall the $B$-spline representation of periodic splines and then derive their matrix formulation in the subspace of periodic splines with zero integral. Building on this representation, we formulate both smoothing spline and penalized spline estimators, together with data-driven selection of the corresponding smoothing parameters.

\subsection{Periodic B-splines}

First, we recall the basic knowledge of the $B$-spline representation of splines; for details, see \cite{deboor78}, \cite{dierckx93}. Let the sequence of knots $\Delta\lambda \, := \, \lambda_{0}=a<\lambda_{1}<\ldots<\lambda_{g}<b=\lambda_{g+1}$  be given. In the following, ${\cal S}_{k}^{\Delta\lambda}[a,b]$ denotes the vector space of polynomial splines of degree $k>0$, defined on a finite interval $[a,b]$ with the sequence of knots $\Delta\lambda$. It is known that $dim\left({\cal S}_{k}^{\Delta\lambda}[a,b]\right)=g+k+1$. Then every spline $s_{k}(x)\in{\cal S}_{k}^{\Delta\lambda}[a,b]$ has a unique representation
\begin{equation*}
    s_{k}\left(x\right) \, = \, \sum_{i=-k}^{g}b_{i}B_{i}^{k+1}\left(x\right).
\end{equation*}
For this representation, it is necessary to add some additional knots, e.g. such that
$$
\lambda_{-k}\, \leq \, \cdots \, \leq \, \lambda_{-1} \, \leq \, \lambda_{0}, \qquad \lambda_{g+1} \, 
\leq \, \lambda_{g+2} \, \leq \, \cdots \, \leq \, \lambda_{g+k+1}.
$$
The vector ${\bf b}=(b_{-k},\ldots,b_{g})^{\top}$ is called {\it the vector of the $B$-spline coefficients} of~$s_{k}(x)$, the functions $B_{i}^{k+1}\left(x\right)$, $i=-k,\ldots,g$ are {\it $B$-splines of degree $k$} and form the basis in ${\cal S}_{k}^{\Delta\lambda}[a,b]$. In matrix notation, the spline $s_{k}(x)\in{\cal S}_{k}^{\Delta\lambda}[a,b]$ can be written as
\begin{equation*}
    s_{k}\left(x\right) \, = \, \mathbf{C}_{k+1}(x)\mathbf{b},
\end{equation*}
where 
$\mathbf{C}_{k+1}(x) \, = \, \left(B_{-k}^{k+1}(x),\dots,B_{g}^{k+1}(x)\right)$ is a vector  of $B$-spline functions. 
% and vector $\mathbf{b} \, = \, \left(b_{-k},\dots,b_{g}\right)^{\top}$ is a vector of $B$-spline coefficients.

A spline function $s_k(x)\in {\cal S}_{k}^{\Delta\lambda}[a,b]$ is called {\it periodic,} if it satisfies periodic boundary conditions
\begin{equation}\label{per_BC}
    s_k^{(l)}(a) \; = \; s_k^{(l)}(b) \qquad \forall \; l \, = \, 0,1,,\ldots, k-1.
\end{equation}
For a spline $s_k(x)\in {\cal S}_{k}^{\Delta\lambda}[a,b]$ to ensure the fulfillment of the periodic boundary conditions~(\ref{per_BC}) it is necessary to add {\em periodic boundary knots} for $i \, = \, 1, \cdots, k$ in the form of \cite{dierckx93},
\begin{align*}
    \lambda_{-i} \; &  = \; \lambda_{g+1-i}\, - \, b \, + \, a,\\
    \lambda_{i+g+1} \; & = \; \lambda_{i} \, + \, b \, - \, a.
\end{align*}
This choice of additional knots is suitable for imposing periodicity conditions~(\ref{per_BC}). With this choice, it follows that 
\begin{equation*}
    B_{i}^{k+1} (x) \; \equiv \; B_{g+1-i}^{k+1}(x+b-a), \qquad i \, = \, 1, \cdots, k.
\end{equation*}
Consequently, taking into account the properties of the $B$-splines, condition~(\ref{per_BC}) will be satisfied if and only if
\begin{equation*}
    b_{-i} \; = \; b_{g+1-i}, \qquad i \, = \, 1, \cdots, k.
\end{equation*}
It is obvious that if we denote
$$
\mathcal{S}_{k,per}^{\Delta\lambda}[a,b] \; = \; 
\left\{s_k(x) \in \mathcal{S}_{k}^{\Delta\lambda}[a,b]: \; s_k^{(l)}(a) \; = \; s_k^{(l)}(b) \; \;\forall \; l \, = \, 0,1,,\ldots, k-1\right\},
$$
then 
$$
dim\left({\cal S}_{k,per}^{\Delta\lambda}[a,b]\right) \; = \; g+1.
$$
Thus, every spline $s_k(x)\in \mathcal{S}_{k,per}^{\Delta\lambda}[a,b]$ has a unique representation
\begin{equation*}
    s_{k}\left(x\right) \, = \, \sum_{i=-k}^{g-k}b_{i}\widetilde{B}_{i}^{k+1}\left(x\right),
\end{equation*}
where
\begin{equation}\label{Bspline_tilde}
    \widetilde{B}_{i}^{k+1}(x) \; = \; 
    \begin{cases}
        B_{i}^{k+1}(x) & \mbox{for} \; i \, = \, 0,\cdots, g-k, \\ 
        B_{i}^{k+1}(x) \, + \, B_{g+1+i}^{k+1}(x) & \mbox{for} \; i \, = \, -k,\cdots, -1.
    \end{cases}
\end{equation}
In matrix notation, the spline 
$s_{k}(x)\in{\cal S}_{k,per}^{\Delta\lambda}[a,b]$ can be written as
\begin{equation}\label{spline_matrix}
    s_{k}\left(x\right) \, = \, \widetilde{\mathbf{C}}_{k+1}(x)\widetilde{\mathbf{b}},
\end{equation}
where 
$\widetilde{\mathbf{C}}_{k+1}(x) \, = \, \left(\widetilde{B}_{-k}^{k+1}(x),\dots, \widetilde{B}_{g-k}^{k+1}(x)\right)$ is a vector of $B$-spline functions $\widetilde{B}_{i}^{k+1}(x)$, $i=-k,\ldots, g-k$, defined in (\ref{Bspline_tilde}) and  vector $\widetilde{\mathbf{b}} \, = \, \left(b_{-k},\dots,b_{g-k}\right)^{\top}$ is a vector of corresponding $B$-spline coefficients. For working with software R or Matlab, it will be definitely useful to write a spline $s_{k}(x)\in{\cal S}_{k,per}^{\Delta\lambda}[a,b]$ by using classical $B$-splines. By the definition of the matrix
\begin{equation}\label{matrix_K}
    \mathbf{K} \; = \; \left(
    \begin{array}{ll}
        \mathbf{I}_k         & \mathbf{0}_{k, g-k+1}\\
        \mathbf{0}_{g-k+1,k} & \mathbf{I}_{g-k+1}\\
        \mathbf{I}_k         & \mathbf{0}_{k, g-k+1}
    \end{array}\right)
\end{equation}
we can write $\widetilde{\mathbf{C}}_{k+1}(x)=\mathbf{C}_{k+1}(x)\mathbf{K}$ and for $s_{k}(x)\in{\cal S}_{k,per}^{\Delta\lambda}[a,b]$ from~(\ref{spline_matrix}) we obtain
\begin{equation}\label{spline_matrix_1}
    s_{k}\left(x\right) \, = \, \widetilde{\mathbf{C}}_{k+1}(x)\widetilde{\mathbf{b}} \, = \, 
    \mathbf{C}_{k+1}(x)\,\mathbf{K}\,\widetilde{\mathbf{b}}.
\end{equation}
If we want to work with periodic splines in $L_0^2[a,b]$ space, it means that we consider subspace
$\mathcal{Z}_{k,per}^{\Delta\lambda}[a,b]$ of space $\mathcal{S}_{k,per}^{\Delta\lambda}[a,b]$ such that 
$$
\mathcal{Z}_{k,per}^{\Delta\lambda}[a,b] \; = \; 
\{s_k(x) \in \mathcal{S}_{k,per}^{\Delta\lambda}[a,b]: \; \int\limits_a^b s_k(x)\,\mbox{d}x \; = \; 0 \}.
$$
It is known \citep{dierckx93} that for $s_k(x)\in \mathcal{S}_{k,per}^{\Delta\lambda}[a,b]$ it holds
$$
\int\limits_a^b s_k(x)\,\mbox{d}x \; = \; \dfrac{1}{k+1} \, \sum\limits_{i=-k}^{g-k} b_i(\lambda_{i+k+1}-\lambda_{i}). 
$$
From this, we can write that $s_k(x)\in \mathcal{S}_{k,per}^{\Delta\lambda}[a,b]$ fulfills the condition
$$
\int\limits_a^b s_k(x)\,\mbox{d}x \; = \; 0  
$$
if and only if,
\begin{equation}\label{coef}
    b_{g-k} \; = \; -\dfrac{1}{\lambda_{g+1}-\lambda_{g-k}} \, \sum\limits_{i=-k}^{g-k-1}b_i(\lambda_{i+k+1}-\lambda_{i}).    
\end{equation}
So one can see that 
$$
dim\left({\cal Z}_{k,per}^{\Delta\lambda}[a,b]\right) \; = \; g.
$$
For the representation of the spline $s_k(x)\in \mathcal{Z}_{k,per}^{\Delta\lambda}[a,b]$, we define the matrix
\begin{equation}\label{matrix_P}
    \mathbf{P} \; = \; \left(
    \begin{array}{c}
        \mathbf{I}_g \\
        \mathbf{a} 
    \end{array}\right),
\end{equation}
where
$$
\mathbf{a} \, = \, -\dfrac{1}{\lambda_{g+1}-\lambda_{g-k}}\left(\lambda_1-\lambda_{-k},\cdots,
\lambda_g-\lambda_{g-k-1}\right).
$$
Thus, by $\mathbf{P}\overline{\mathbf{b}}=\widetilde{\mathbf{b}}$ we have a vector of $B$-spline coefficients that satisfy condition (\ref{coef}). With respect to this and (\ref{spline_matrix_1}), the spline $s_k(x)\in \mathcal{Z}_{k,per}^{\Delta\lambda}[a,b]$
can be written
\begin{equation}\label{spline_matrix_per_zero}
    s_{k}\left(x\right) \, = \, \widetilde{\mathbf{C}}_{k+1}(x) \, \mathbf{P}\, \overline{\mathbf{b}} \, = \,     \mathbf{C}_{k+1}(x)\,\mathbf{K}\, \mathbf{P} \, \overline{\mathbf{b}},
\end{equation}
where $\overline{\mathbf{b}} \, = \, (b_{-k},\cdots, b_{g-k-1})^{\top}$. We used the idea that for periodic spline with zero integral on interval $[a,b]$ we have
$$
\widetilde{\mathbf{b}} \, = \, \mathbf{P}\,\overline{\mathbf{b}}.
$$
Finally, if we set
$
\mathbf{b} \, = \, \mathbf{K}\,\mathbf{P}\,\overline{\mathbf{b}},
$
then for every $s_k(x)\in \mathcal{Z}_{k,per}^{\Delta\lambda}[a,b]$ we have representation
\begin{equation*}
s_{k}\left(x\right) \, = \, \mathbf{C}_{k+1}(x) \, \mathbf{b}.
\end{equation*}

%%%%%%%%%%%%%%%%%%%%%%%%%%%%%%%%%%%%%%%%%%%%%%%%%%%%%%%%%%%%%%%%

\subsection{Periodic smoothing splines}

%%%%%%%%%%%%%%%%%%%%%%%%%%%%%%%%%%%%%%%%%%%%%%%%%%%%%%%%%%%%%%%%

Having established the $B$-spline representation of periodic splines, we now focus on the construction of the periodic smoothing spline with zero integral over the interval $I=[a,b]$. In this case, the zero integral constraint must be imposed while maintaining the periodic character of the spline. Following the idea from \cite{machalova16}, where the zero integral condition was incorporated through the spline coefficients in the non-periodic setting, we adopt a similar coefficient-based strategy here. This approach allows us to enforce the constraint directly in the coefficient formulation without modifying the underlying periodic $B$-spline basis. As a result, a compact and computationally efficient matrix representation can be derived for the periodic spline space with zero integral.

In \cite{machalova21}, an alternative concept, the $Z\!B$-spline basis, was introduced. Each $Z\!B$-spline basis function has zero integral over the interval, which makes this basis suitable for non-periodic smoothing splines in $L_0^2(I)$. However, because the construction relies on multiple boundary knots, it cannot satisfy periodic boundary conditions. Therefore, in the periodic case, the coefficient-based formulation described above must be used.

In the following, we derive the matrix formulation of the periodic smoothing spline with zero integral, based on the standard penalized least squares functional. We express the solution in terms of the periodic $B$-spline basis and show how the zero integral constraint can be incorporated directly into the coefficient system.

For this purpose, let data $(x_{i},y_{i})$, $a\leq x_{i}\leq b$, weights $w_{i}>0$, $i=1,\ldots,n$, sequence of knots $\Delta\lambda=\left\{\lambda_i\right\}_{i=0}^{g+1}$, $\lambda_{0}=a<\lambda_{1}<\ldots<\lambda_{g}<b=\lambda_{g+1}$, $n\geq g+1$ and a parameter $\alpha\in(0,1)$ be given. 

Now our task is to find  spline $s_{k}(x)\in{\cal Z}_{k,per}^{\Delta\lambda}[a,b]\subset L_0^2(I)$, which for arbitrary $l\in\left\{1,\ldots,k-1\right\}$ minimizes the functional
\begin{equation*}%\label{jl}
    J_{l}(s_k) \; = \; (1-\alpha)\int_{a}^{b}\left[s_{k}^{(l)}(x)\right]^{2}\, \mbox{d}x \, +  \, 
    \alpha \sum\limits_{i=1}^{n} w_{i}\left[y_{i}-s_{k}(x_{i})\right]^{2}.
\end{equation*}
The choice of parameter $\alpha$ and derivation $l$ affects the smoothness of the resulting spline. Let us denote $\mathbf{x}=\left(x_{1},\ldots,x_{n}\right)^{\top}$, $\mathbf{y}=\left(y_{1},\ldots,y_{n}\right)^{\top}$, $\mathbf{w}=\left(w_{1},\ldots,w_{n}\right)^{\top}$ and $\mathbf{W}=diag\left(\mathbf{w}\right)$. Regarding the representation (\ref{spline_matrix_per_zero}) the functional $J_l(s_k)$ can be written as a quadratic function
\begin{equation}\label{functional}
\begin{split}
  J_l(\,\overline{\mathbf{b}}\,) \; = & \; (1-\alpha)\,{\overline{\bf b}}^{\top}\,{\bf P}^{\top}\,
  {\bf K}^{\top}\,{\bf S}_l^{\top}\,{\bf M}_{kl}\,{\bf S}_l\,{\bf K\,P\,\overline{b}}\, + \\
	& \hspace{1cm} + \,\alpha \left[{\bf y}-{\bf C}_{k+1}({\bf x})\,{\bf K\,P\,\overline{b}}\,\right]^{\top} {\bf W} \left[{\bf y}-{\bf C}_{k+1}({\bf x}){\bf K\,P\,\overline{b}}\,\right],
\end{split}
\end{equation}
where the matrix $\mathbf{K}$, $\mathbf{P}$ is defined by formulae (\ref{matrix_K}) and (\ref{matrix_P}), respectively. Further, the matrix
\begin{equation*}
  {\bf M}_{kl} \; = \; \left( m_{ij}^{kl}\right)_{i,j=-k+l}^{g}, \quad\mbox{with}\quad m_{ij}^{kl} \; = \; \int\limits_{a}^{b}B_{i}^{k+1-l}(x)B_{j}^{k+1-l}(x)\,\mbox{d}x,
\end{equation*}
is positive definite because $B_{i}^{k+1-l}(x)\geq 0$, $i=-k+l,\ldots,g$ are linear independent functions. The upper triangular matrix ${\bf S}_{l}={\bf D}_{l}{\bf L}_{l}\ldots{\bf D}_{1}{\bf L}_{1}\in\mathbb{R}^{g+k+1-l,g+k+1}$ has full row rank, matrix ${\bf D}_{j}\in\mathbb{R}^{g+k+1-j,g+k+1-j}$ is a diagonal matrix such that
$$
{\bf D}_{j}=\left(k+1-j\right)diag\left(d_{-k+j},\ldots,d_{g}\right)
$$
with
$$
d_{i}=\dfrac{1}{\lambda_{i+k+1-j}-\lambda_{i}}, \qquad i=-k+j,\ldots,g,
$$
and
$${\bf L}_{j}:=
\left(\begin{array}{cccc}
      -1 & 1      &        &  \\
         & \ddots & \ddots &  \\
         &        & -1     & 1
      \end{array} 
\right)\in\mathbb{R}^{g+k+1-j,g+k+2-j}.
$$
Finally, ${\bf C}_{k+1}({\bf x})\in\mathbb{R}^{n,g+k+1}$ stands for the collocation matrix, i.e.
$$
{\bf C}_{k+1}({\bf x}) \, = \, \left(B_i^{k+1}(x_j)\right)_{j=1,\, i=-k}^{n,\, g}.
$$
Using the notation $\mathbf{U}:= {\bf K}\,{\bf P}$,
\begin{equation}\label{eq:matrix_G}
  {\bf G} \, := \, {\bf U}^{\top}\left[(1-\alpha){{\bf S}_l}^{\top}{\bf M}_{kl}{\bf S}_l+\alpha\, {\bf C}_{k+1}^{\top}({\bf x}){\bf W}\,{\bf C}_{k+1}({\bf x})\right]{\bf U}
\end{equation}
and
$$
{\bf g} \, := \, \alpha {\bf U}^{\top}{\bf C}_{k+1}^{\top}({\bf x}){\bf Wy},
$$
it is possible to rewrite the quadratic function $J_l(\, \overline{\bf b}\,)$ given in (\ref{functional}) as
\begin{equation}\label{jlb}
  J_l(\,\overline{\bf b}\,) \; = \; {\overline{\bf b}}^{\top}{\bf G}\,\overline{\mathbf{b}}-
  2\overline{{\bf b}}^{\top}{\bf g}+\alpha{\bf y}^{\top}{\bf Wy}.
\end{equation}
Our task is to find a spline $s_{k}(x)\in{\cal Z}_{k,per}^{\Delta\lambda}[a,b]$ that minimizes the functional $J_{l}(s_{k})$. In other words, we want to find a vector $\overline{\mathbf{b}}\in \mathbb{R}^g$ that minimizes the function (\ref{jlb}). It is obvious that this function has just one minimum if and only if the matrix ${\bf G}$ is positive definite (p.d.). From~(\ref{eq:matrix_G}), it can be easily seen that
$$
{\bf G} \quad \mbox{ is p.d.} \quad \Leftrightarrow \quad {\bf C}_{k+1}(\bf x) \quad \mbox{is of full column rank}.
$$
From the Schoenberg-Whitney theorem and its generalization, see \cite{deboor78}, it is known that the matrix ${\bf C}_{k+1}(\bf x)$ is of full column rank if and only if there exists $\{u_{-k},\ldots, u_{g}\}\subset\{x_1,\ldots,x_n\}$ with $u_i<u_{i+1}$, $i=-k,\ldots,g-1$, such that $\lambda_i<u_i<\lambda_{i+k+1}$, $i=-k,\ldots,g$. In this case, from the necessary and sufficient condition for a unique minimum of a quadratic function, i.e.
$$
\frac{\partial J_{l}(\, \overline{\bf b}\,)}{\partial\overline{\bf b}^{\top}} \, = \, 0,
$$
we get a system of linear equations ${\bf G}\overline{\bf b} \, = \, {\bf g}$ and then the unique solution of this system is given by
\begin{equation}\label{b*}
    \overline{\bf b}^{*} \, = \, {\bf G}^{-1}{\bf g}.
\end{equation}
The resulting smoothing spline is obtained using the formula
\begin{equation*}
    s_{k}^{*}(x) \, = \, \sum\limits_{i=-k}^{g-k-1}b_{i}^{*}\widetilde{B}_{i}^{k+1}(x),
\end{equation*}
in matrix notation, using standard $B$-splines $B_i^{k+1}(x)$ as
\begin{equation*}\label{szm*}
  s_{k}^{*}(x) \, = \, \mathbf{C}_{k+1}(x)\mathbf{KP}\overline{\mathbf{b}}^*,
\end{equation*}
where the vector $\overline{\mathbf{b}}^{*}=\left(b_{-k}^{*},\ldots,b_{g-k-1}^{*}\right)^{\top}$ is given in (\ref{b*}).

The smoothing parameter $\alpha$ in \eqref{functional} controls the trade-off between fidelity to the data and the smoothness of the resulting spline. Therefore, an appropriate choice of $\alpha$ is essential to obtain a well-balanced periodic smoothing spline. A commonly used approach for its selection is the \textit{generalized cross-validation} (GCV) criterion, originally introduced by \cite{craven1979}.

Let $\mathbf{H}(\alpha)$ denote the projection (hat) matrix associated with the smoothing spline $s_{k}^{*}(x)$, i.e.
\begin{equation*}
  \mathbf{H}(\alpha) = \alpha\, \mathbf{C}_{k+1}(\mathbf{x})\, \mathbf{U} \mathbf{G}^{-1} \mathbf{U}^{\top} \mathbf{C}_{k+1}^{\top}(\mathbf{x}) \mathbf{W},
\end{equation*}
where $\mathbf{G}$ is the matrix defined in~\eqref{eq:matrix_G}. Then the GCV criterion is defined as
\begin{equation}\label{eq:GCV}
  \mathrm{GCV}(\alpha)
  = \frac{1}{n}\,
    \frac{\sum_{i=1}^{n}\left(y_i - s_k^*(x_i)\right)^{2}}
         {\bigl(1 - \operatorname{trace}(\mathbf{H}(\alpha))/n\bigr)^{2}}.
\end{equation}
The optimal value of the smoothing parameter is obtained as:
\[
  \alpha_{\mathrm{opt}} = \arg\min_{\alpha \in (0,1)} \mathrm{GCV}(\alpha).
\]
As noted in \cite{wahba1990}, the GCV criterion provides a stable and computationally efficient data-driven way to determine the smoothing parameter while avoiding the tendency of standard cross-validation to under-smooth the solution.

An alternative and widely used approach to smoothing is based on the concept of \emph{penalized splines} (P-splines), see \cite{eilers2021}. In this case, the roughness penalty is not formulated through the derivatives of the spline function but rather through finite differences of the spline coefficients. 
Let $\mathbf{D}_d$ denote the $d$-th order difference matrix then the penalized least squares functional for a periodic spline with zero integral can be written as
\begin{equation}\label{eq:P_spline_functional}
  J(\,{\overline{\mathbf{b}}}\,) \;=\;
  (\mathbf{y}-\mathbf{C}_{k+1}(\mathbf{x})\mathbf{KP}\overline{\mathbf{b}})^{\top}\mathbf{W}(\mathbf{y}-\mathbf{C}_{k+1}(\mathbf{x})\mathbf{KP}\overline{\mathbf{b}})
  + \rho \, \overline{\mathbf{b}}^{\top}\mathbf{D}_d^{\top}\mathbf{D}_d \overline{\mathbf{b}},
\end{equation}
where $\rho>0$ is a penalization parameter. Similarly to the minimization of~\eqref{jlb} we can find a minimum of \eqref{eq:P_spline_functional} that is given by
\begin{equation*}
    \overline{\bf b}^{*}_P \, = \, {\bf G}_P^{-1}{\bf g}_P, 
\end{equation*}
where
\[
{\bf G}_P \, = \, {\bf U}^{\top}\left[ {\bf C}_{k+1}^{\top}({\bf x}){\bf W}\,{\bf C}_{k+1}({\bf x}) + \rho\, \mathbf{D}_d^{\top}\mathbf{D}_d\right]{\bf U}
\]
and
\[
{\bf g}_P \, = \, {\bf U}^{\top}{\bf C}_{k+1}^{\top}({\bf x}){\bf Wy}.
\]
The selection of the penalization parameter $\rho$ can be carried out in the same manner as before, using the GCV criterion \eqref{eq:GCV}, where only the corresponding hat matrix $\mathbf{H}_P(\rho)$ is in the form
\[
\mathbf{H}_P(\rho) = \mathbf{C}_{k+1}(\mathbf{x})\mathbf{U} \mathbf{G}_P^{-1}\mathbf{U}^{\top} \mathbf{C}_{k+1}^{\top}(\mathbf{x})\, \mathbf{W}
\]
is used in place of $\mathbf{H}(\alpha)$. So this gives us the formula.
\begin{equation}\label{eq:GCV_P}
  \mathrm{GCV}(\rho)
  = \frac{1}{n}\,
    \frac{\sum_{i=1}^{n}\left(y_i - s_k^*(x_i)\right)^{2}}
         {\bigl(1 - \operatorname{trace}(\mathbf{H}_P(\rho))/n\bigr)^{2}}.
\end{equation}
This approach provides a practical alternative to derivative-based smoothing, as it avoids computing integrals of squared derivatives and thus simplifies the numerical implementation while remaining directly applicable to the periodic setting with the zero integral constraint.
\section{Application to wind direction data}
\label{sec:5}

Environmental datasets were collected at Pardubice Airport, located in the central part of the Czech Republic (50°00'N, 15°44'E and 230 m above sea level). Data were manually downloaded on a monthly basis from the meteomanz.com website. The measurements span the period from January 1, 2014, to December 31, 2023, with a temporal resolution of one hour, resulting in approximately 87,500 records. The dataset consists of measurements of temperature, relative humidity, pressure, wind direction, wind speed, and cloudiness.

\begin{figure}[htbp]
    \centering
    \includegraphics[width=2.6in]{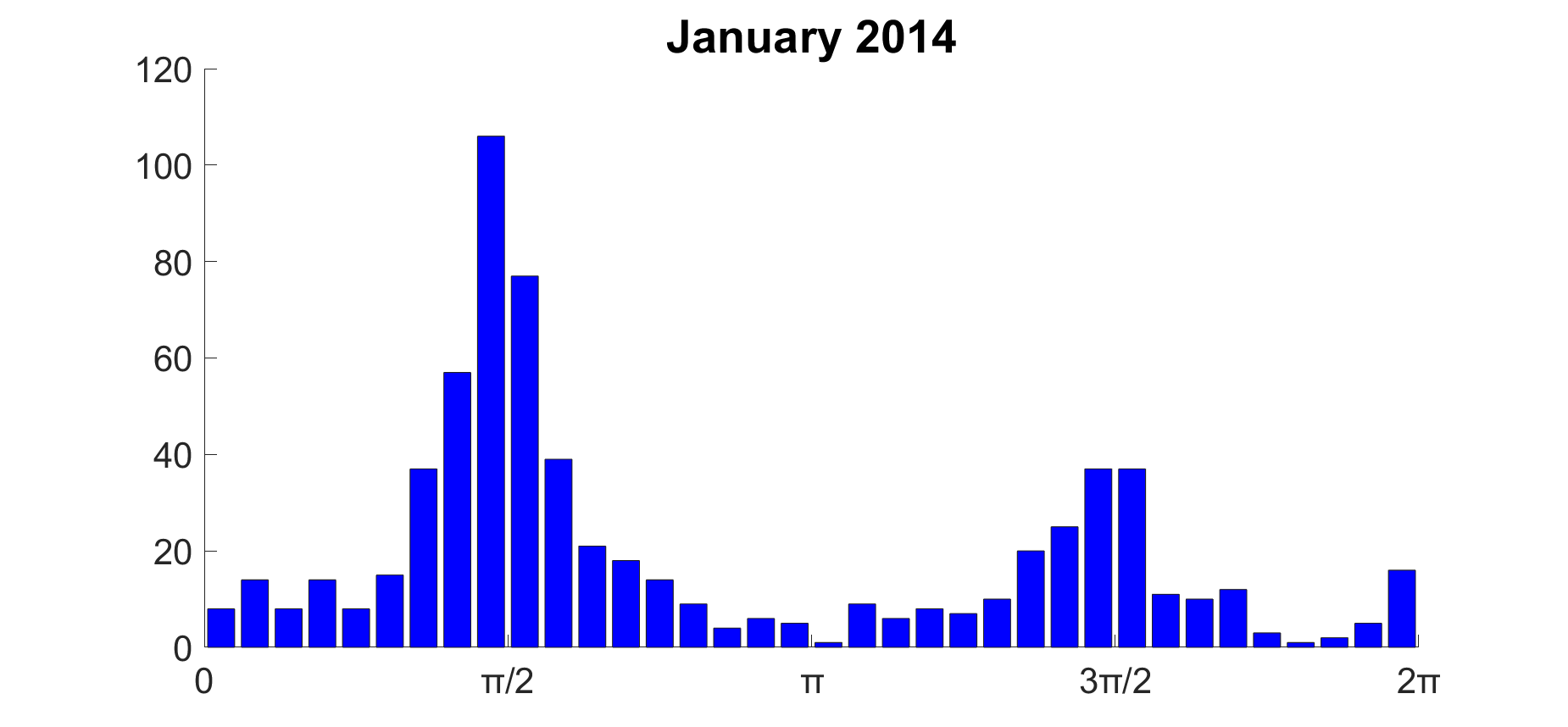}
    \includegraphics[width=2.6in]{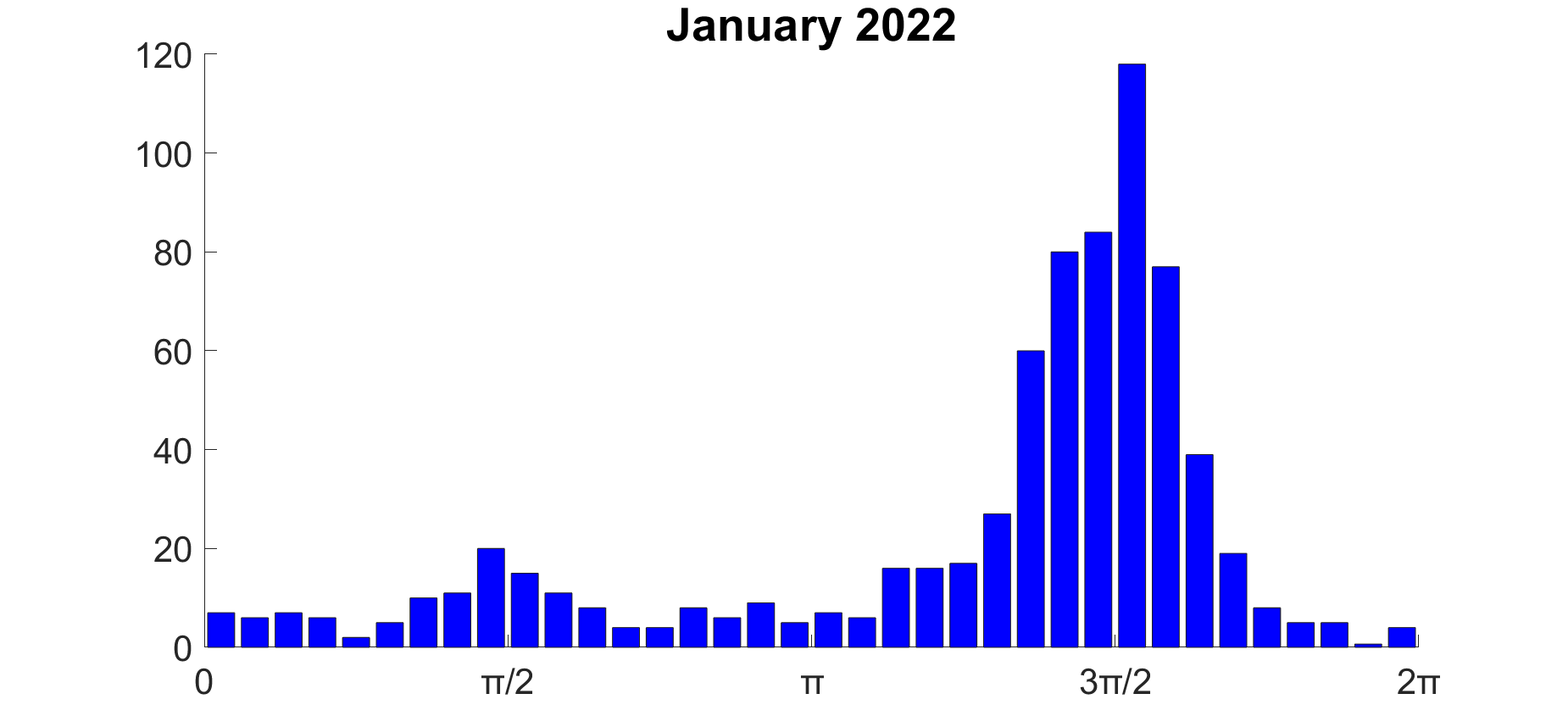}
    \caption{Histogram for January in the years 2014 and 2022}
    \label{fig:obrazek1}
\end{figure}

\begin{figure}[htbp]
    \centering
    \includegraphics[width=0.45\textwidth]{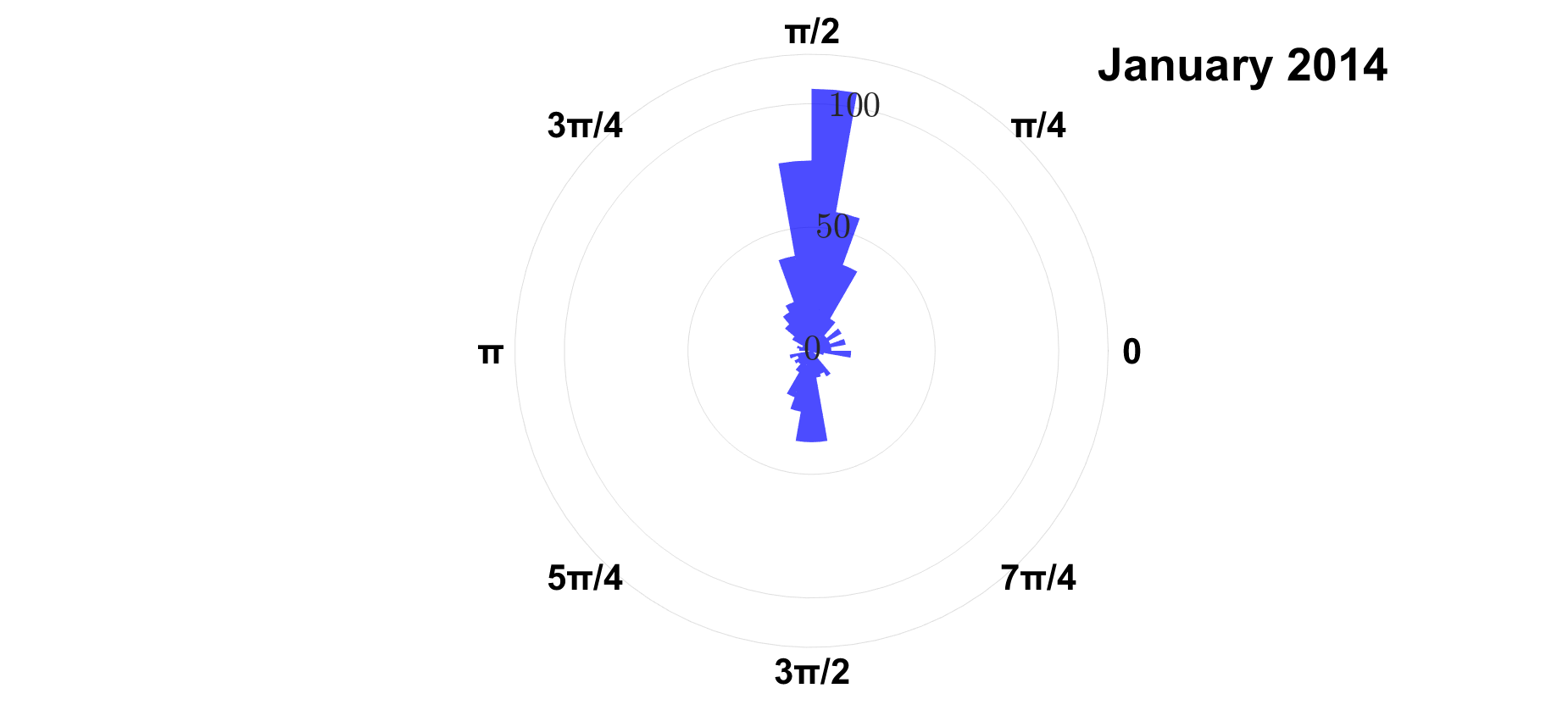}
    \includegraphics[width=0.45\textwidth]{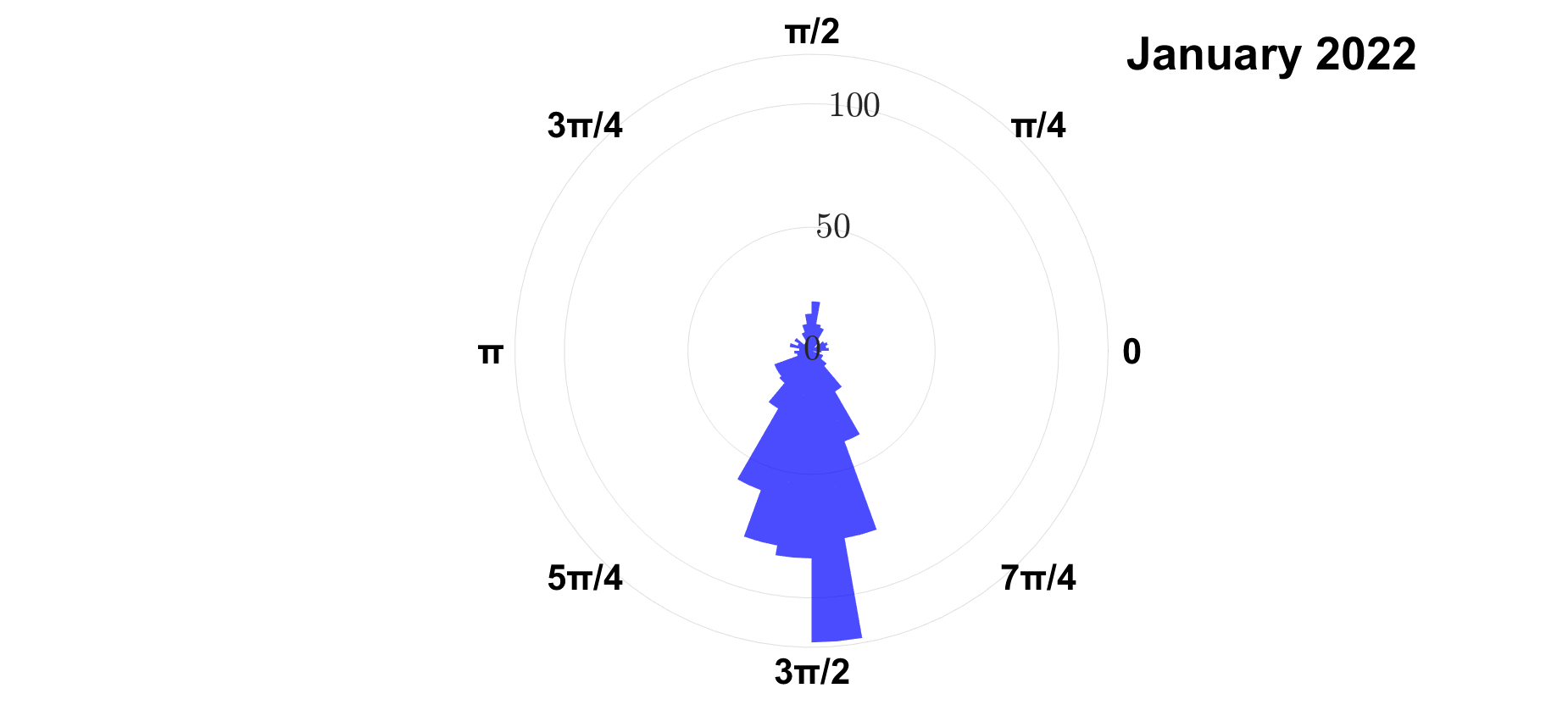}
      \caption{Wind rose for January in the years 2014 and 2022}
        \label{fig:obrazek2} 
\end{figure}

To obtain pure wind direction datasets for each year, the original measurements required preprocessing. Specifically, calm wind records (zero speed and undefined direction) were excluded, and wind direction values were converted from degrees to radians. The resulting datasets can be represented using a histogram or a wind rose of absolute frequencies, as illustrated in Figures~\ref{fig:obrazek1} and~\ref{fig:obrazek2}.

\subsection{Periodic spline approximation}

To numerically and graphically illustrate the construction of periodic spline approximations of wind direction distributions, we employed datasets covering the period 2014--2023. Each dataset corresponds to a single month and consists of hourly measurements collected over all days of that month. For each monthly dataset, the histogram of wind direction was first transformed into relative frequencies in order to approximate the probability density function. These densities were then mapped from the Bayes space $\mathcal{B}^2(I)$ to the Hilbert space $L_0^2(I)$ by the discrete version of centered log-ratio (clr) transformation. All smoothing procedures described below were performed in the clr-transformed space $L_0^2(I)$, which ensures compatibility with the standard functional data analysis setting. After obtaining the spline approximation, the resulting functions were back-transformed by the inverse clr transformation to recover the estimated densities in $\mathcal{B}^2(I)$.
 
For the construction of smoothing splines or P-splines, nine inner equidistant knots were used,
\[
\Delta \lambda := \{\lambda_i\}_{i=0}^{10}, \quad \lambda_i = \pi i / 5, \quad i = 0,1,\dots,10.
\]
We considered four variants of periodic splines, differing in the type and order of the smoothness penalty:

\begin{itemize}
    \item[(a)] Smoothing spline: \(k = 3,\; l = 1\) \\
    First-order derivative penalty in the functional \(J_l\), providing flexible fits that follow local variations in the data.

    \item[(b)] Smoothing spline: \(k = 3,\; l = 2\) \\
    Second-order derivative penalty, yielding smoother periodic curves and reducing small oscillations.

    \item[(c)] P-spline: \(k = 3,\; d = 1\) \\
    Penalization based on first-order finite differences of spline coefficients, resulting in moderate smoothness and computational efficiency.

    \item[(d)] P-spline: \(k = 3,\; d = 2\) \\
    Penalization through second-order finite differences, producing the smoothest approximations with stable behaviour across all datasets.
\end{itemize}

For each of these four variants, the optimal value of the smoothing or penalization parameter, i.e. \(\alpha\) for the smoothing splines and \(\rho\) for the P-splines, was determined by means of the generalized cross-validation (GCV) criterion defined in~(\ref{eq:GCV}) for the smoothing splines and analogously in~(\ref{eq:GCV_P}) for the P-splines. The optimization was performed separately for every dataset, i.e., for each month of every year in the period 2014--2023. For the smoothing spline variants, the mean optimal parameters obtained across all datasets were approximately \(\alpha_{\text{opt}}^{(a)} = 0.927\) for \(l = 1\) and \(\alpha_{\text{opt}}^{(b)} = 0.990\) for \(l = 2\). For the P-spline variants, the corresponding averaged penalization parameters were \(\rho_{\text{opt}}^{(c)} = 0.070\) for \(d = 1\) and \(\rho_{\text{opt}}^{(d)} = 0.041\) for \(d = 2\). These values were then used for the subsequent construction of the final periodic smoothing and P-splines.

All four variants (a)--(d) share the same squared term of the objective functional
\begin{equation*}
\text{SSE}(s_k) \, = \, \sum_{i=1}^{n}(y_i - s_k(x_i))^2,
\end{equation*}
which quantifies the deviation between the spline and the observed data. The difference among the variants lies only in the specific type of smoothing and penalization, while the squared term remains identical. The largest and smallest values of SSE for all four variants are given in Table~\ref{tab:sse-extrema}. The largest values were consistently obtained for March~2018 and the smallest for September~2023.

\begin{table}[H]
\begin{center}
\caption{Months with maximum and minimum SSE values for each variant (a)--(d).}
\label{tab:sse-extrema}
{\small{\tabcolsep 6pt\renewcommand{\arraystretch}{1.25}\small
\begin{tabular}{|c|c|c|}
\hline
\textbf{Variant} & \textbf{Max SSE} & \textbf{Min SSE} \\
\hline
(a) & 3 676 & 327 \\
(b) & 3 992 & 330 \\
(c) & 4 072 & 355 \\
(d) & 4 447 & 377 \\
\hline
\end{tabular}}}
\end{center}
\end{table}

To provide a more global comparison of model performance, Table~\ref{tab:sse-mean} summarizes the average values of the squared term SSE computed over all available months for each variant~(a)--(d). These mean values reflect the overall goodness of fit and allow for a quantitative comparison between the smoothing spline and P-spline formulations. Although the absolute differences are relatively small, the ranking of the variants remains consistent with the qualitative behaviour observed in the graphical results.

\begin{table}[H]
\begin{center}
\caption{Average values of SSE over all months for each variant.}
\label{tab:sse-mean}
{\small
\tabcolsep 6pt
\renewcommand{\arraystretch}{1.25}
\begin{tabular}{|c|c|c|c|c|}
\hline
\textbf{Variant} & \textbf{(a)} & \textbf{(b)} & \textbf{(c)} & \textbf{(d)} \\
\hline
\textbf{Mean SSE} & 1 102 & 1 167 & 1 183 & 1 263 \\
\hline
\end{tabular}
}
\end{center}
\end{table}

\begin{figure}[htbp]
    \centering
    \includegraphics[width=0.49\textwidth]{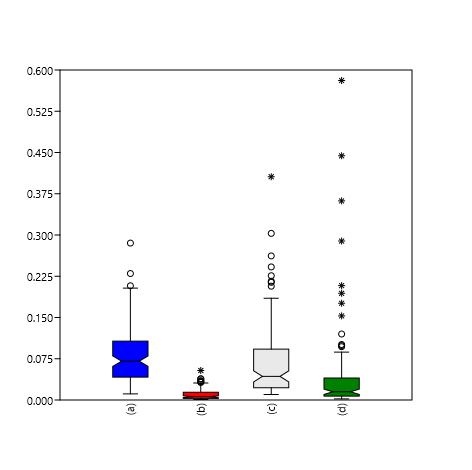}
    \hfill
    \includegraphics[width=0.49\textwidth]{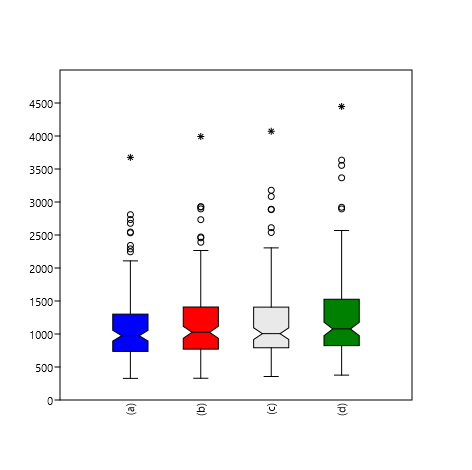}
    \caption{ Boxplots of smoothing or penalization parameter (left) and SSE values (right) for approximation variants (a)-(d)}
        \label{fig:4V_violin} 
\end{figure}

Visualization by boxplots can be used to compare the distribution of the smoothing parameter \(\alpha\), the penalization parameter \(\rho\), and the SSE values of the objective functional for the individual variants, as illustrated in Figure~\ref{fig:4V_violin}. The smoothing parameter \(\alpha\) affects both parts of the objective functional~(\ref{functional}) and forms their convex combination. In contrast, the penalization parameter \(\rho\) appears only in the second part of the functional~(\ref{eq:P_spline_functional}) through the difference matrix. Therefore, in order to compare the smoothing parameter \(\alpha\) and the penalization parameter \(\rho\), we quantify the smoothing effect as $ \frac{ ( 1 - \alpha )} {\alpha} $, which is directly comparable to \(\rho\). The left panel of Figure~\ref{fig:4V_violin} illustrates the distribution of the smoothing effect for variants (a) and (b) and the distribution of penalization for the P-spline variants (c) and (d). It can be seen that increasing $l$ and $d$ for variants (b) and (d) results in a lower smoothing or penalization effect. In contrast to smoothing splines, whose values are more concentrated around the centre, P-spline approximations exhibit many outliers. The right panel of Figure~\ref{fig:4V_violin} compares the distribution of SSE values for the selected approximation variants. Variant (a), the smoothing spline with $k=3$ and $l=1$, has the smallest variation range and average value. For the other approximation variants (b)--(d), these characteristics gradually increase, although the differences in the distributions are not substantial.  

\begin{figure}%[htbp]
    \centering
    \includegraphics[width=0.49\textwidth]{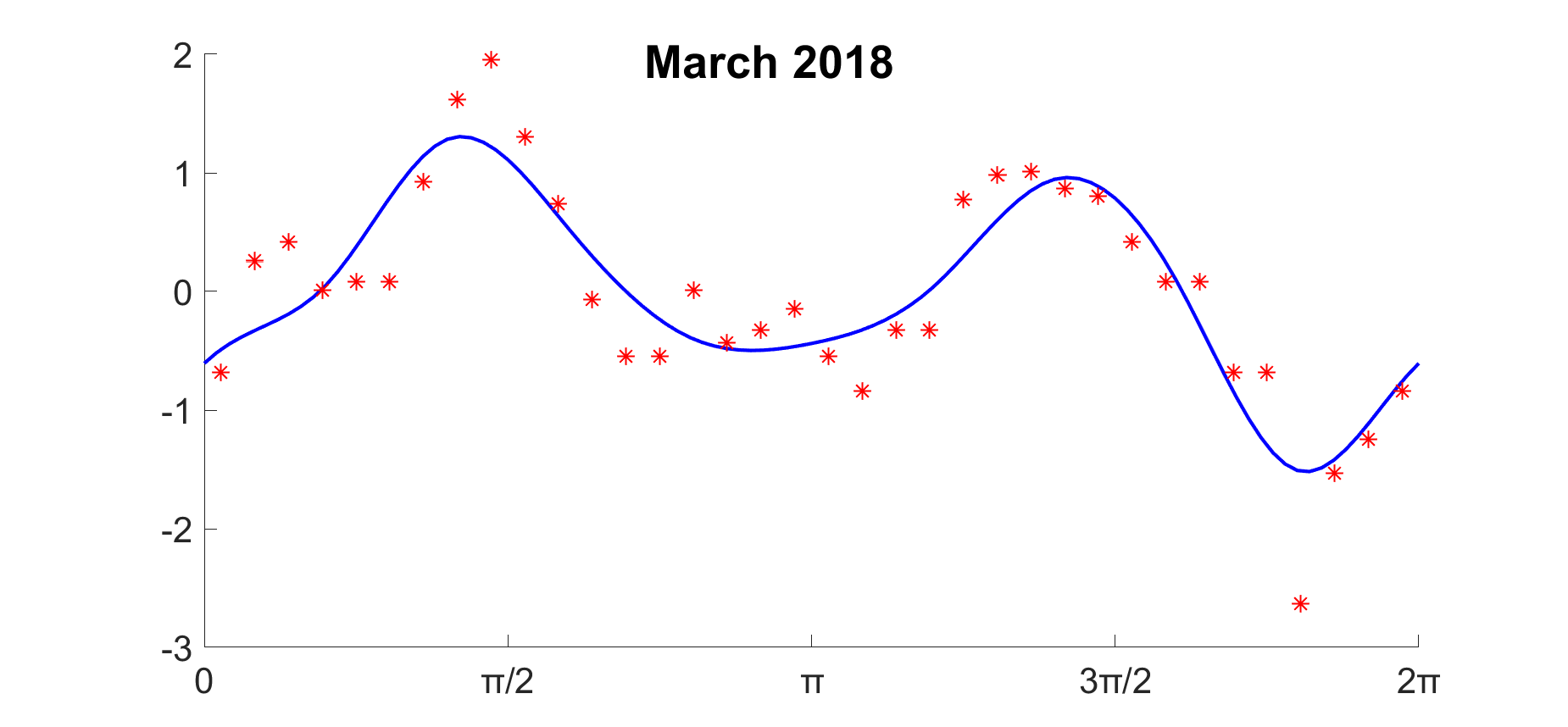}
    \hfill
    \includegraphics[width=0.49\textwidth]{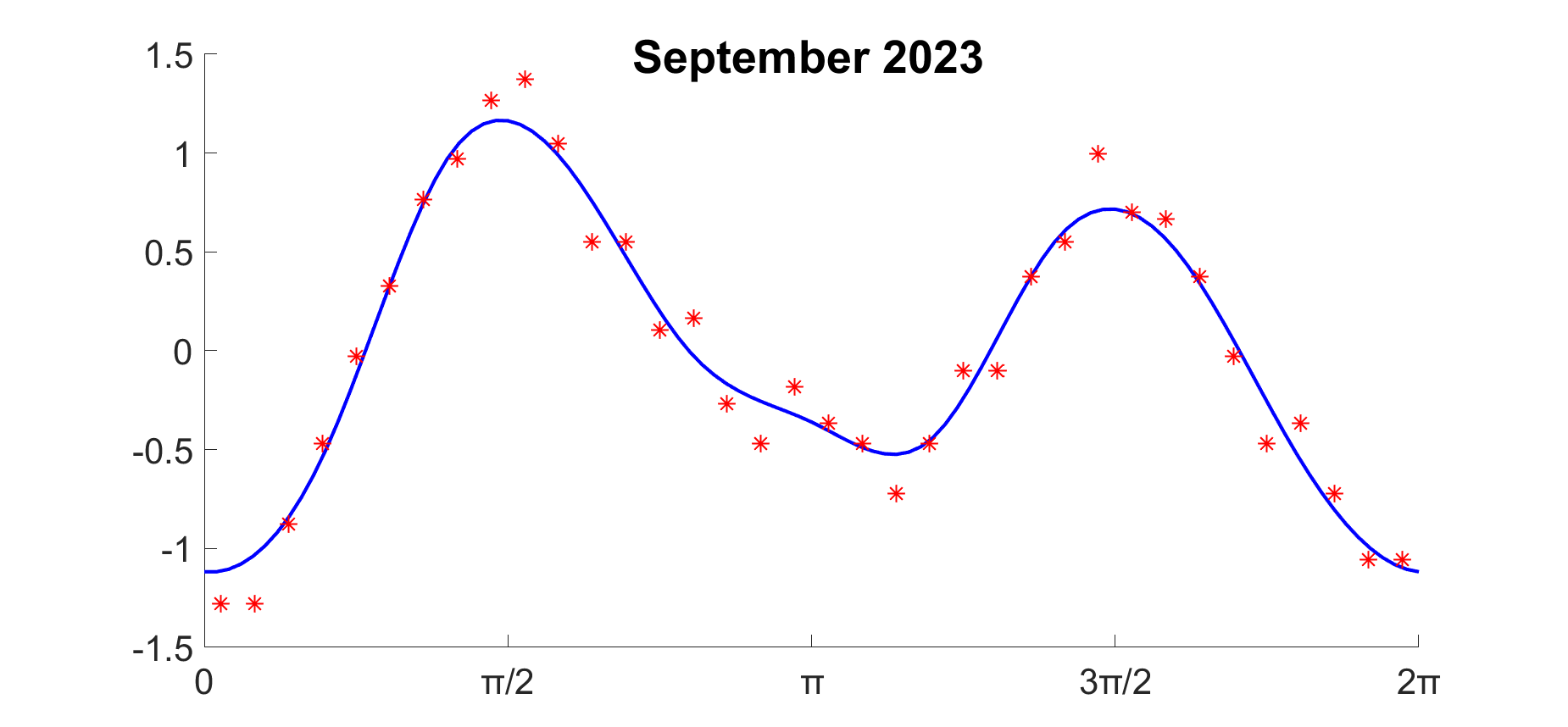}
    \caption{Splines in $L_0^2(0,2\pi)$ for March 2018 (left) and September 2023 (right)}
    \label{fig:clr_splines} 
\end{figure}

\begin{figure}%[h]%[htbp]
    \centering
    \includegraphics[width=0.49\textwidth]{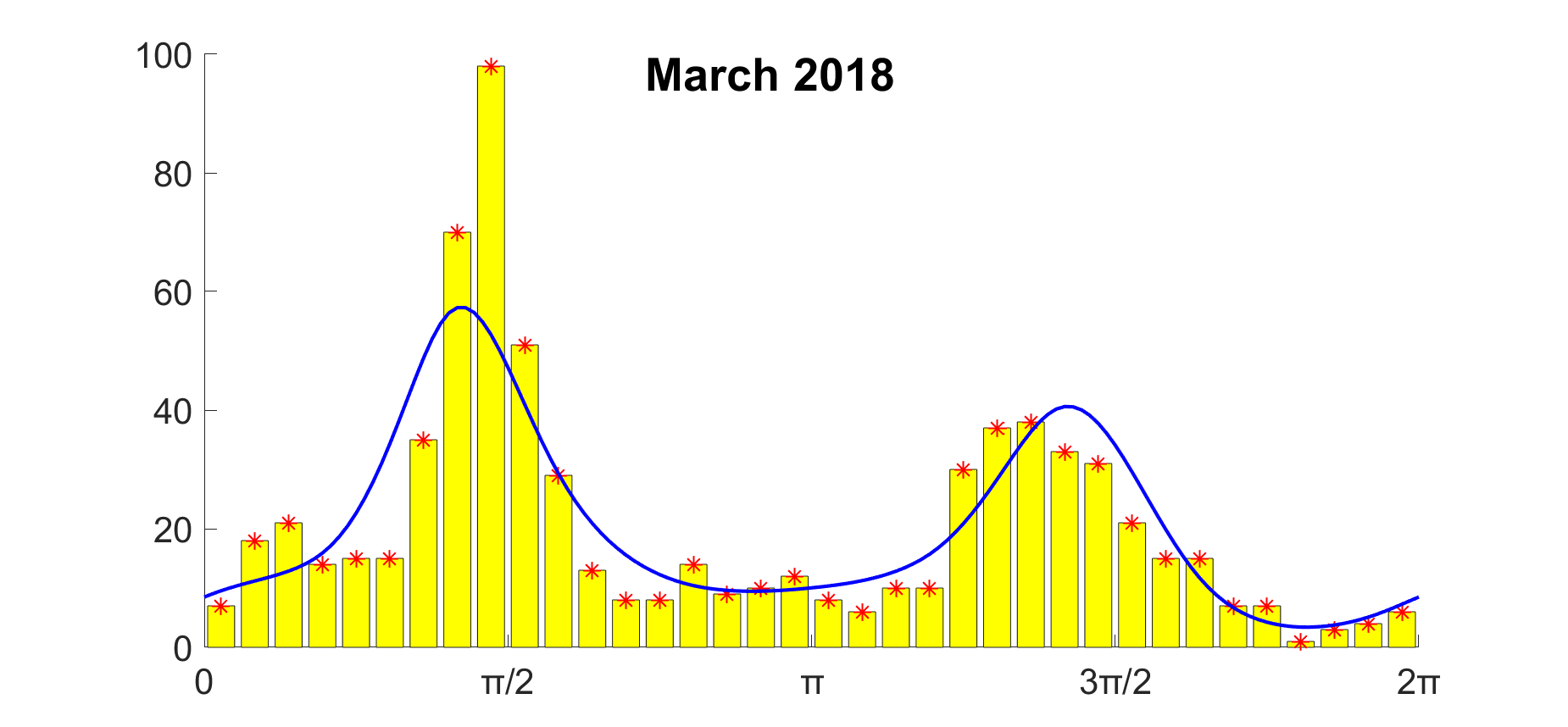}
    \hfill
    \includegraphics[width=0.49\textwidth]{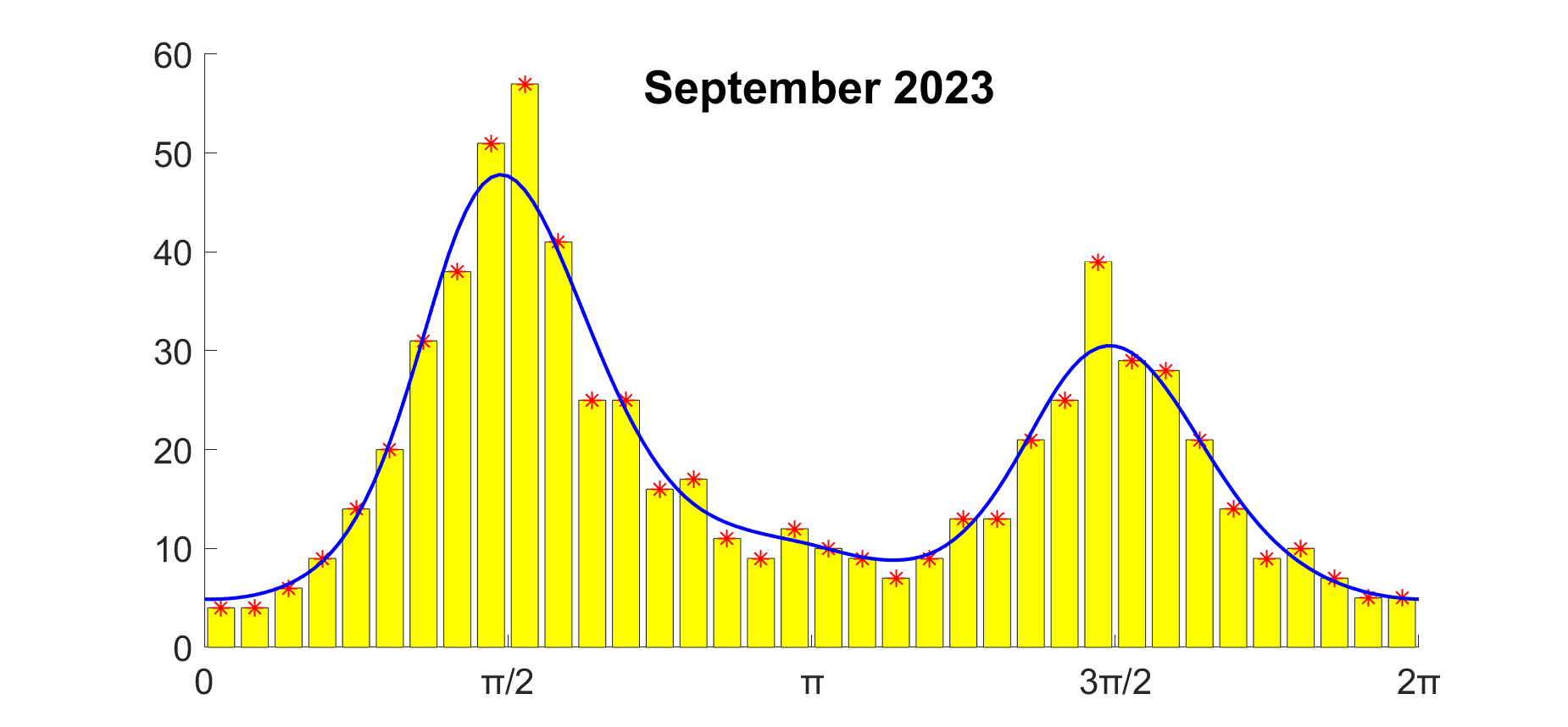}
       \caption{Cubic spline in $\mathcal{B}^2(0,2\pi)$ for March 2018 (left) and September 2023 (right)}
        \label{fig:iclr_splines} 
\end{figure}

Figures~\ref{fig:clr_splines} and~\ref{fig:iclr_splines} provide representative examples for March~2018 and September~2023, corresponding to the largest and smallest values of the SSE among all months and all variants, respectively. In Figure~\ref{fig:clr_splines}, the periodic spline is shown in the clr domain $L_0^2(0,2\pi)$, where the zero-integral constraint is imposed and the smoothing is performed. Figure~\ref{fig:iclr_splines} then displays the original histogram together with its spline approximation, allowing a direct visual assessment of the fit on the original data scale. In Figure~\ref{fig:iclr_splines} on the left, it may appear that the extreme value at $\pi/2$ is not estimated accurately. Nevertheless, it is generally known that the inverse clr transformation shifts the variability towards lower values, which visually affects the result in $\mathcal{B}^2(0,2\pi)$. 

\begin{figure}[h]
    \centering
    \includegraphics[width=0.49\textwidth]{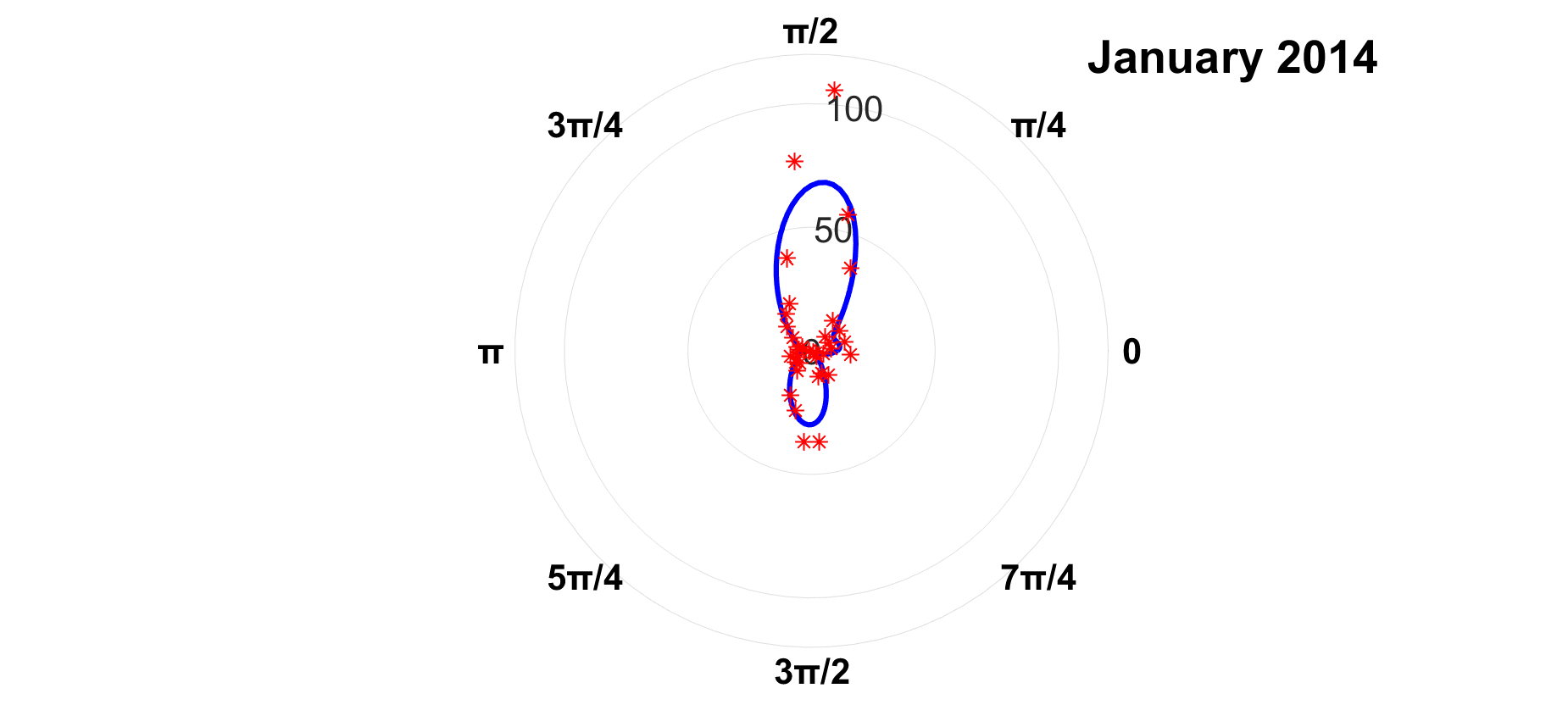}
    \hfill
    \includegraphics[width=0.49\textwidth]{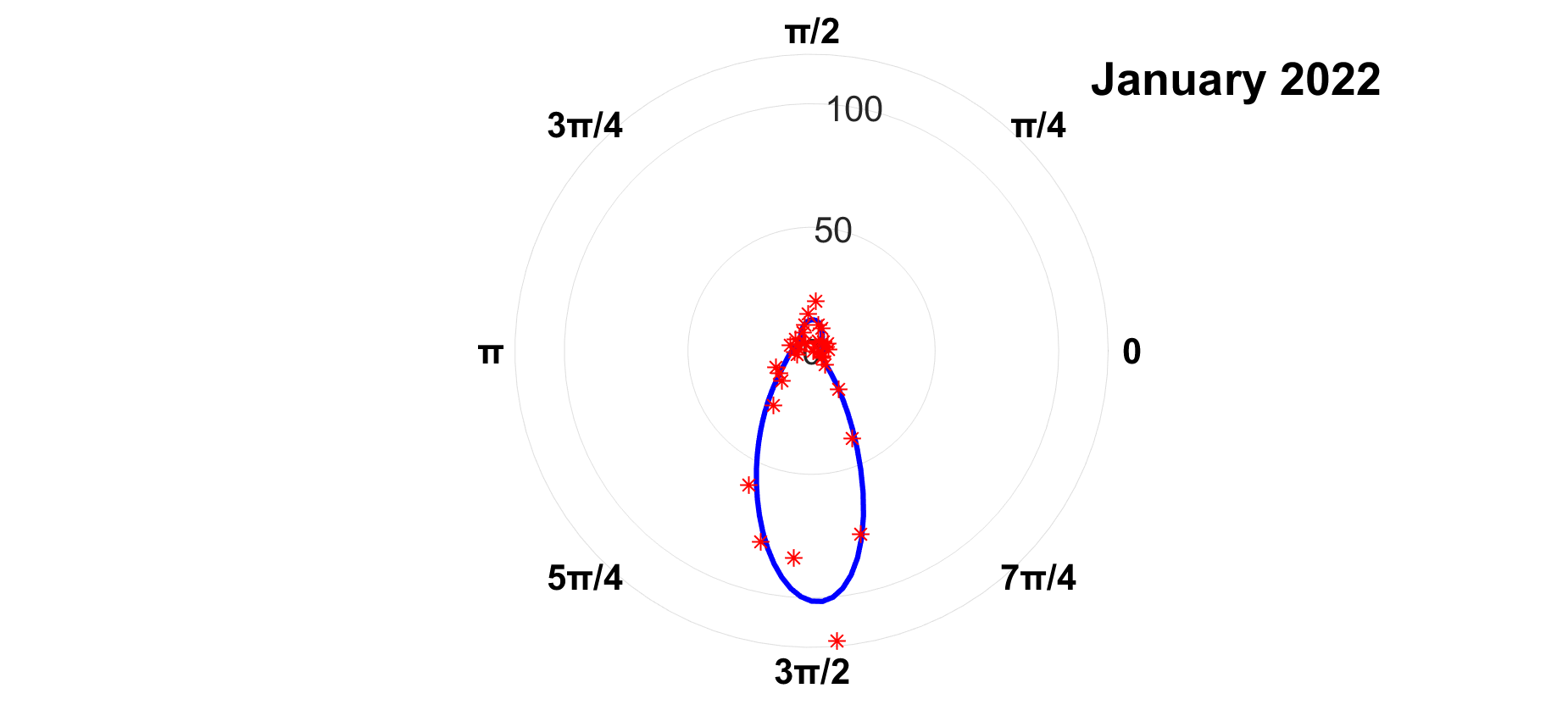}
      \caption{Splines in polar coordinates $\mathcal{B}^2(0,2\pi)$ for January 2014 (left) and 2022 (right)}
        \label{fig:polar_splines} 
\end{figure}

Sometimes, it may be useful to plot the smoothing splines in polar coordinates. For example, for January 2014 and 2022, the splines are shown in Figure~\ref{fig:polar_splines}.

\begin{figure}[htbp]
    \centering
    \includegraphics[width=0.49\textwidth]{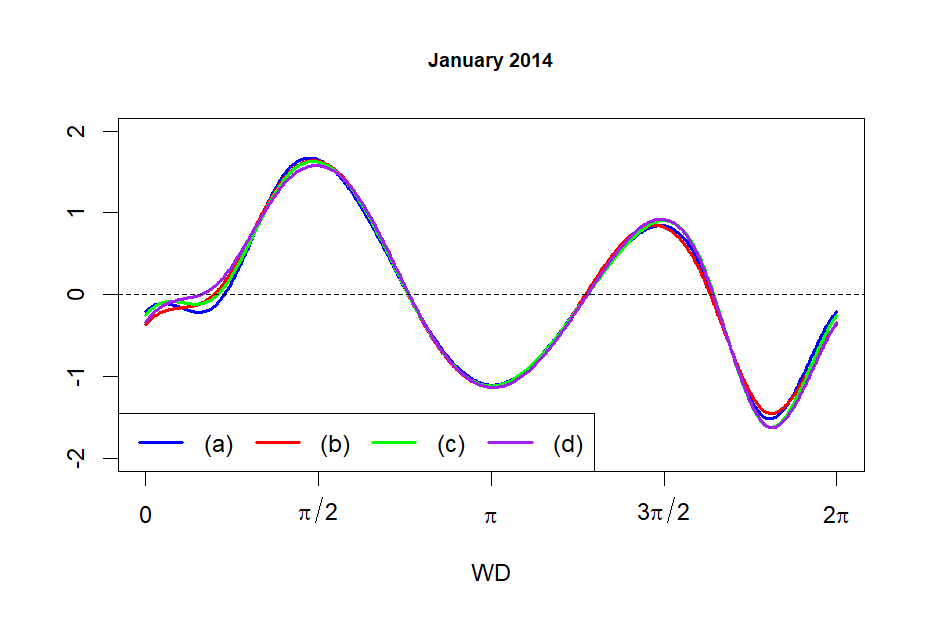}
    \hfill
    \includegraphics[width=0.49\textwidth]{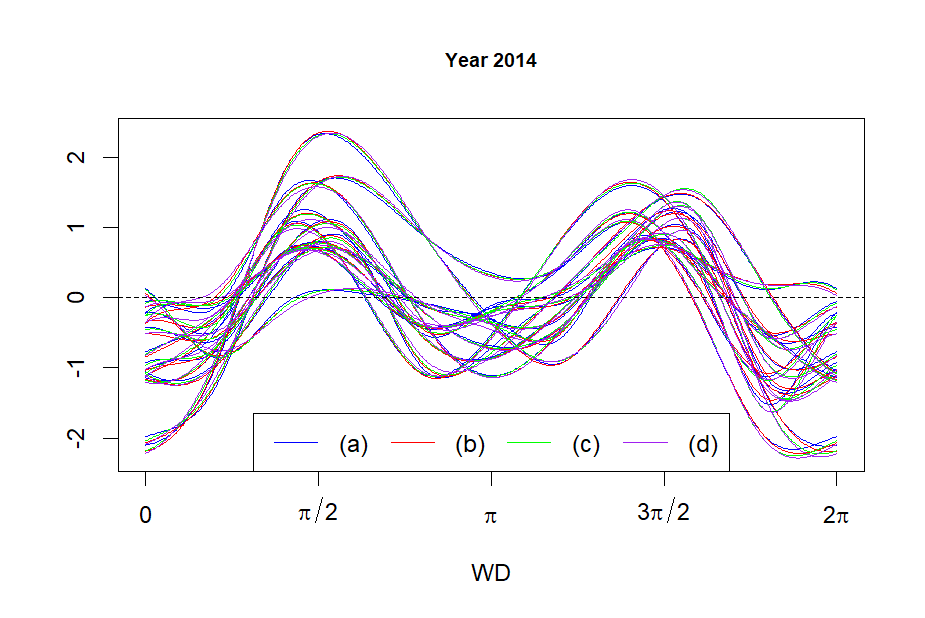}
      \caption{All four variants of splines in $L_0^2(0,2\pi)$ for January 2014 (left) and whole year 2014 (right)}
        \label{fig:4V_M1} 
\end{figure}

Another way to compare the differences between spline variants is through graphical output. Figure~\ref{fig:4V_M1} on the left shows all four variants (a) to (d) of the spline for January 2014 in the space $L_0^2(0,2\pi)$ with zero integral. On the right-hand side of Figure~\ref{fig:4V_M1}, 12 curves for the four variants are shown for the period January--December 2014. It is clear that the differences between variants (a) to (d) are negligible in relation to the variability of the data.

\subsection{Statistical analysis of wind direction dataset}

\begin{figure}[htbp]
    \centering
    \includegraphics[width=0.9\textwidth]{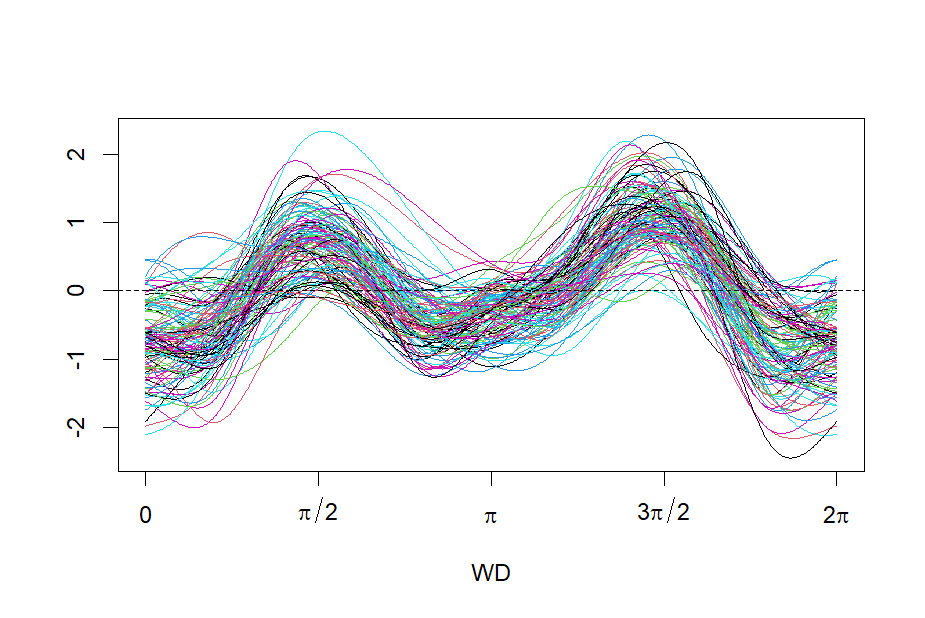}
      \caption{ Dataset of 120 periodic smoothing splines (January 2014 - December 2023) in the space $L_0^2(0,2\pi)$}
        \label{fig:Der1_all} 
\end{figure}

Let us display all 120 curves (10 years with monthly resolution) in the space $L_0^2(0,2\pi)$ for variant (a), which was selected as the best according to the average value of the SSE criterion. Figure~\ref{fig:Der1_all} shows them over the support domain, where all curves are periodic and satisfy the zero-integral constraint.

\begin{figure}[htbp]
    \centering
    \includegraphics[width=0.49\textwidth]{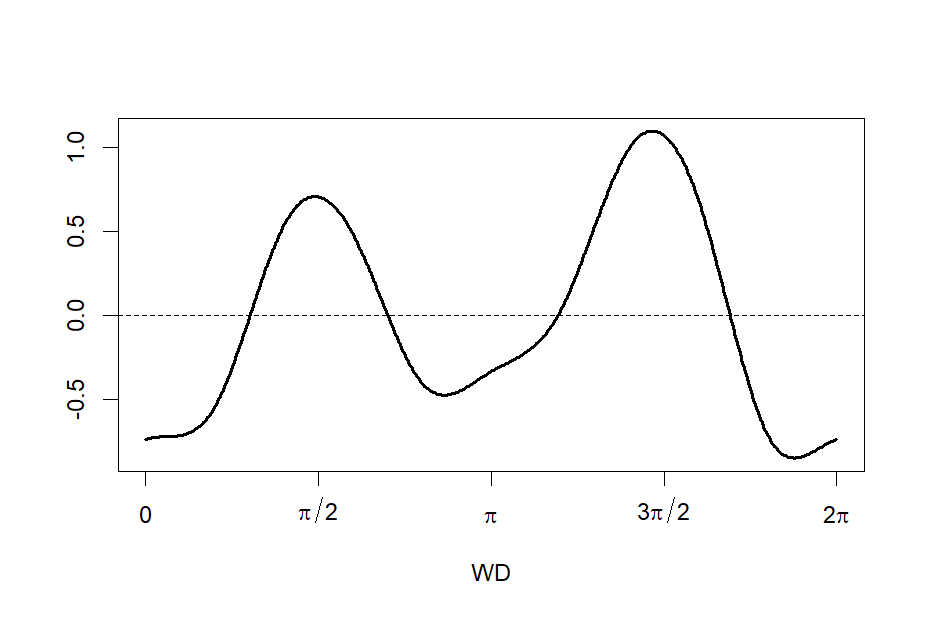}
    \hfill
    \includegraphics[width=0.49\textwidth]
    {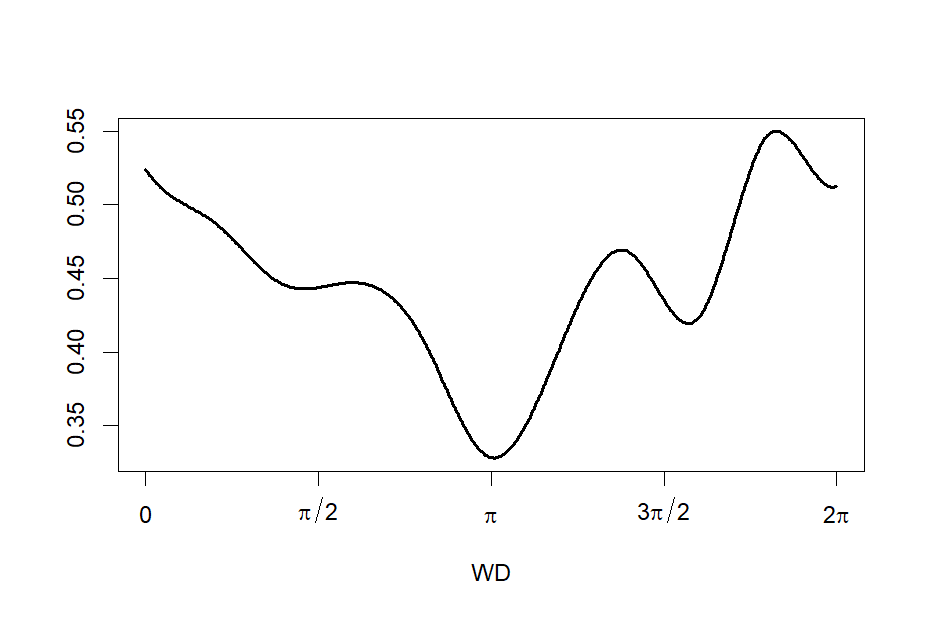}\\
    \includegraphics[width=0.49\textwidth]{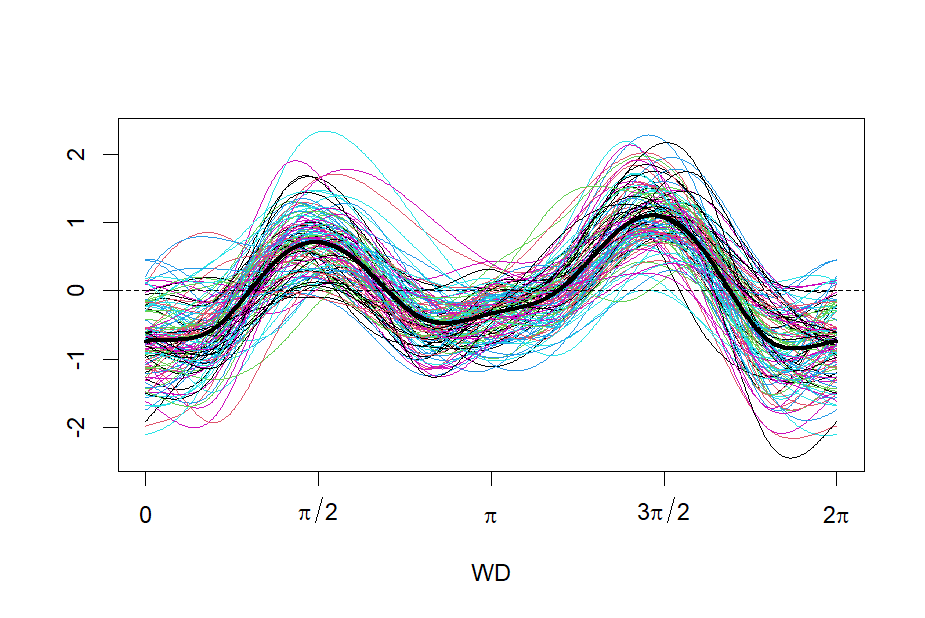}
    \hfill
    \includegraphics[width=0.49\textwidth]{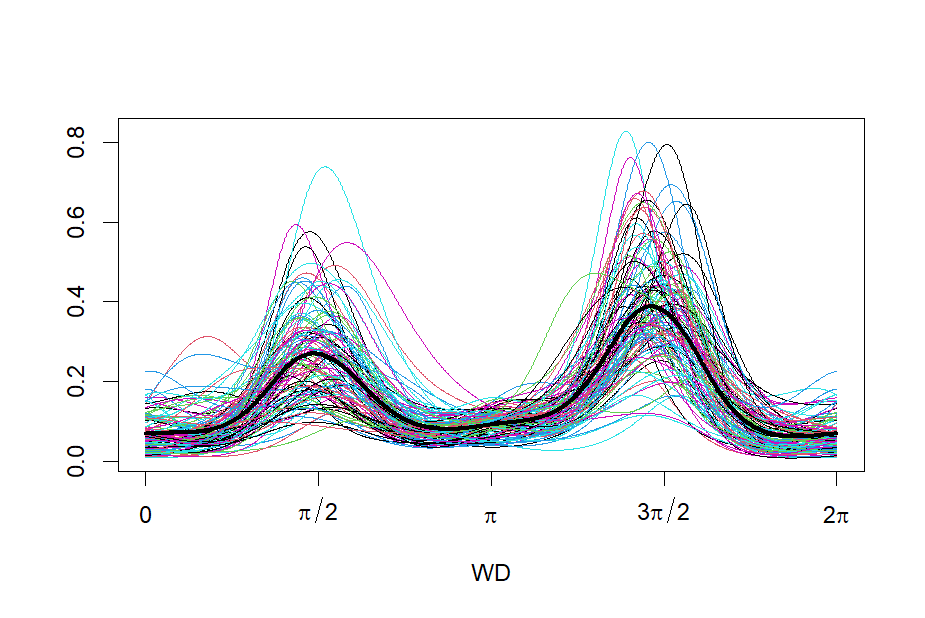}
      \caption{ Average and standard deviation curves of dataset (upper left and right); dataset augmented by average in $L_0^2(0,2\pi)$ resp. $\mathcal{B}^2(0,2\pi)$ (bottom left resp. right)}
        \label{fig:4V_mean} 
\end{figure}

Since we have a dataset of curves in the space $L_0^2(0,2\pi)$, we can apply the FDA methods described in Section~\ref{sec:2} for further statistical processing. In the upper left panel of Figure~\ref{fig:4V_mean}, the sample mean function is shown, while the upper right panel displays the standard deviation function of the dataset. In the lower left panel, all curves are visualized together with the average, and in the lower right panel, the densities after inverse clr transformation are displayed, including the average wind direction density. The annual wind direction distribution at Pardubice Airport is bimodal, with prevailing west-east and east-west directions. The highest variability is concentrated around the north-south direction.

The dependence of wind direction distribution on time is examined using FDA regression, where the explanatory variable is the month index $i=1, \dotsc,120$. Let us build a function-on-scalar regression model from Section \ref{subsec:funcreg} in $L_0^2(0,2\pi)$:
\begin{equation}\label{regT_modelL}
\mathrm{clr} (y_i (t))  \, = \, \mathrm{clr} ( \beta_0(t)) \, + \, \mathrm{clr} ( \beta_1(t)) \cdot i \, + \, \xi_i (t) \quad  i=1, \dotsc,120,\; t \in (0,2\pi),
\end{equation}
where $\mathrm{clr} (y_i (t))$ are the periodic spline representations of wind direction, $\mathrm{clr} ( \beta_0(t))$ is the spline representation of the intercept, $\mathrm{clr} ( \beta_1(t))$ is the spline representation of the time parameter, and $\xi_i (t)$ are i.i.d. random functional errors with zero mean in $L_0^2(0,2\pi)$ that are independent of the time predictor.

After the inverse clr transformation, we obtain the corresponding model in $\mathcal{B}^2(0,2\pi)$:
\begin{equation}\label{regT_modelB}
y_i (t)  \, = \, \beta_0(t) \, \oplus \, [i \odot \beta_1](t) \, \oplus \, \varepsilon_i (t) \quad  i=1, \dotsc,120 , \quad t \in (0,2\pi),
\end{equation}
\noindent
where $y_i (t)$ are wind direction densities, $\beta_0(t)$ and $ \beta_1(t)$ are time regression parameters, and $\varepsilon_i (t) = \mathrm{clr}^{-1} (\xi_i (t))$ are random errors whose mean is the neutral element of perturbation in $\mathcal{B}^2(0,2\pi)$.

\begin{figure}[htbp]
    \centering
    \includegraphics[width=0.45\textwidth]{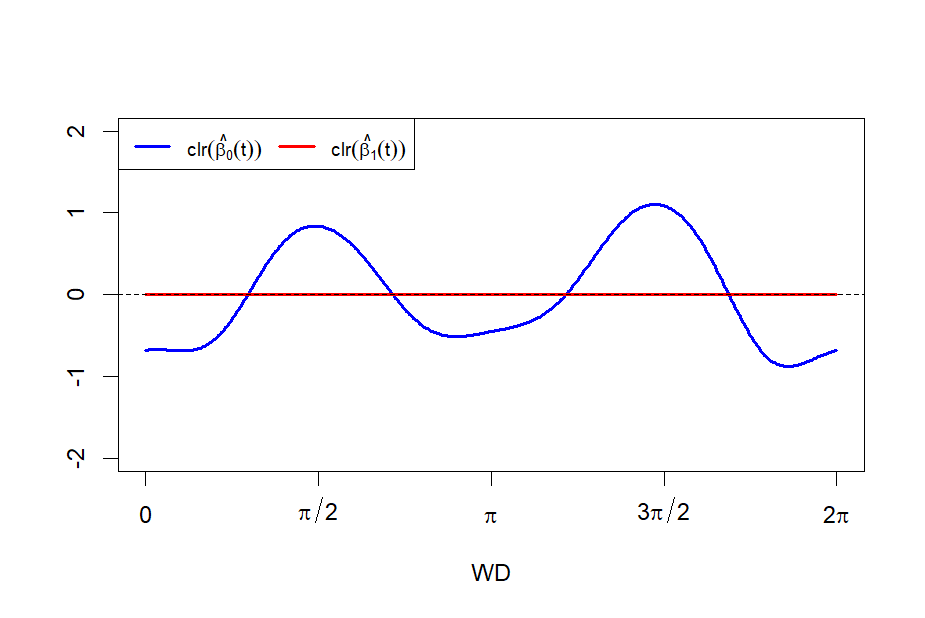}
    \hfill
    \includegraphics[width=0.45\textwidth]{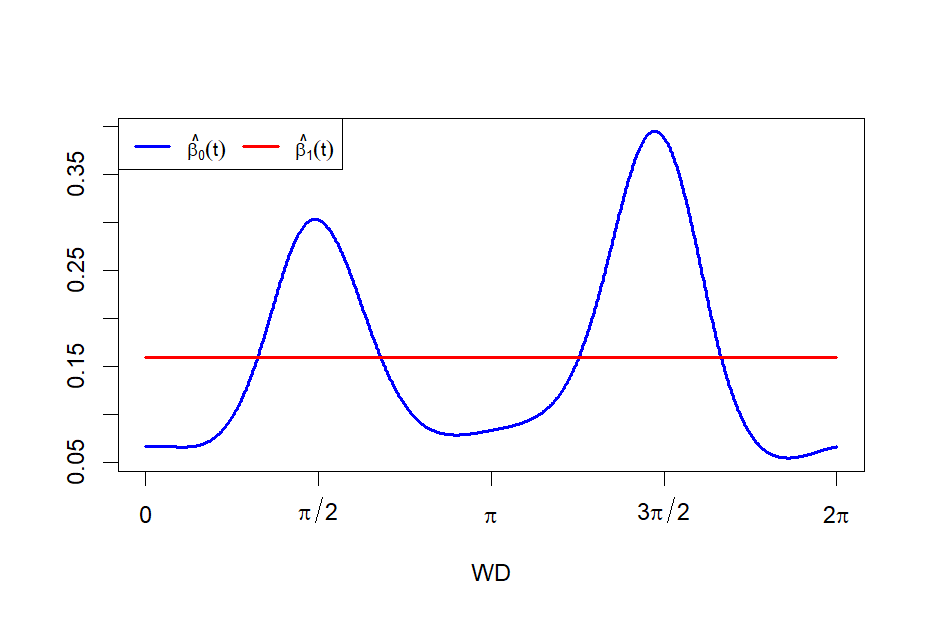} \\
    \includegraphics[width=0.45\textwidth]{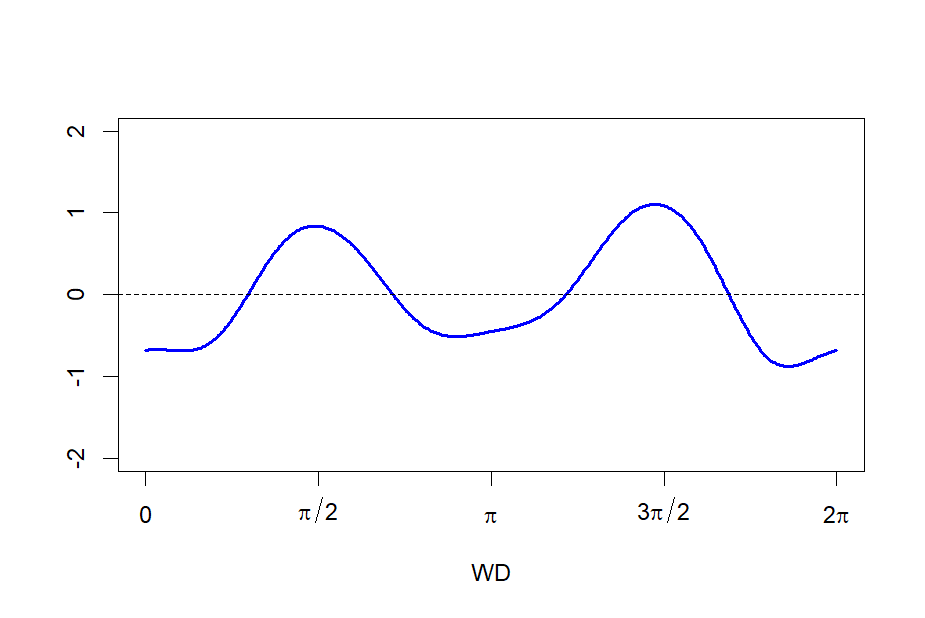}
    \hfill
    \includegraphics[width=0.45\textwidth]{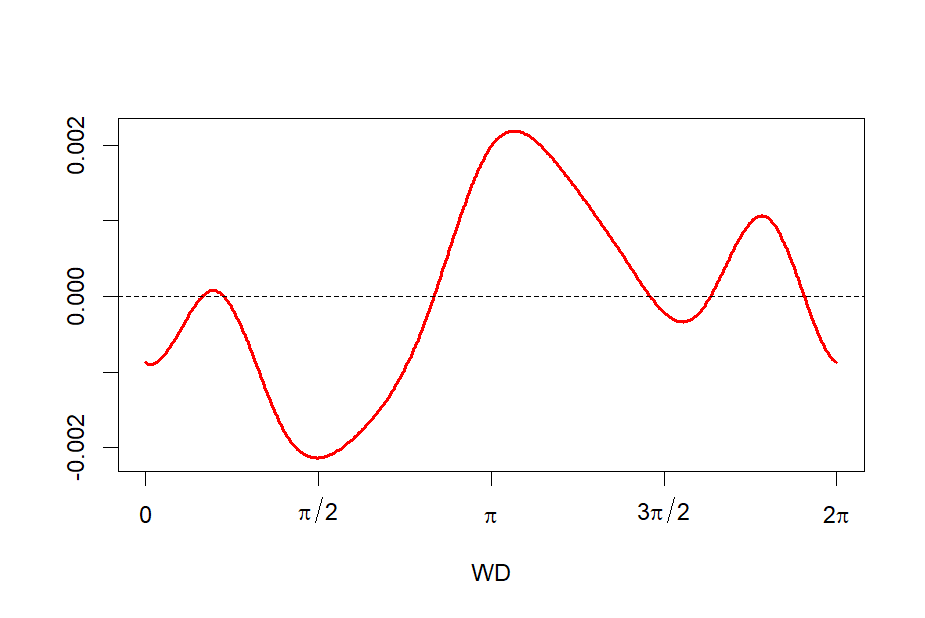} 
      \caption{ Functional time regression parameters in $L_0^2(0,2\pi)$ resp. $\mathcal{B}^2(0,2\pi)$ (upper left resp. right); intercept $\mathrm{clr} (\hat{ \beta_0}(t))$ in $L_0^2(0,2\pi)$ (bottom left) and time parameter $\mathrm{clr} (\hat{ \beta_1}(t))$ in $L_0^2(0,2\pi)$ (bottom right)}
      \label{fig:regT_coef} 
\end{figure}

Figure~\ref{fig:regT_coef} shows both functional regression parameters in $L_0^2(0,2\pi)$ in the upper left panel and the corresponding densities in $\mathcal{B}^2(0,2\pi)$ after inverse clr transformation in the upper right panel. The lower panels visualize the individual shapes of the intercept $\mathrm{clr} (\hat{ \beta_0}(t))$ and the time parameter $\mathrm{clr} (\hat{ \beta_1}(t))$ in $L_0^2(0,2\pi)$. Clearly, the functional slope parameter only deviates negligibly from the neutral element, i.e., uniform density, on the common scale.

\begin{figure}[htbp]
    \centering
    \includegraphics[width=0.45\textwidth]{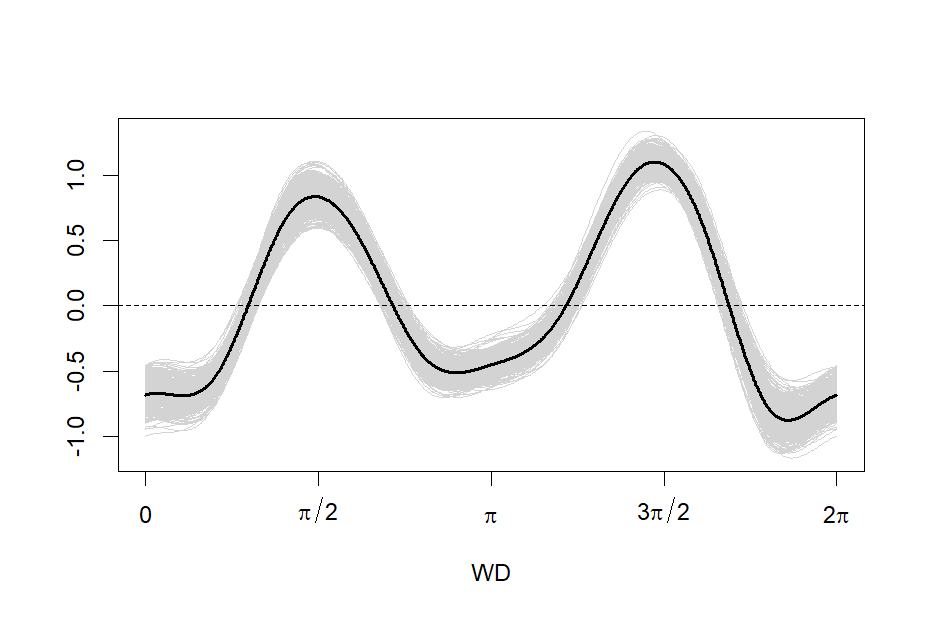}
    \hfill
    \includegraphics[width=0.45\textwidth]{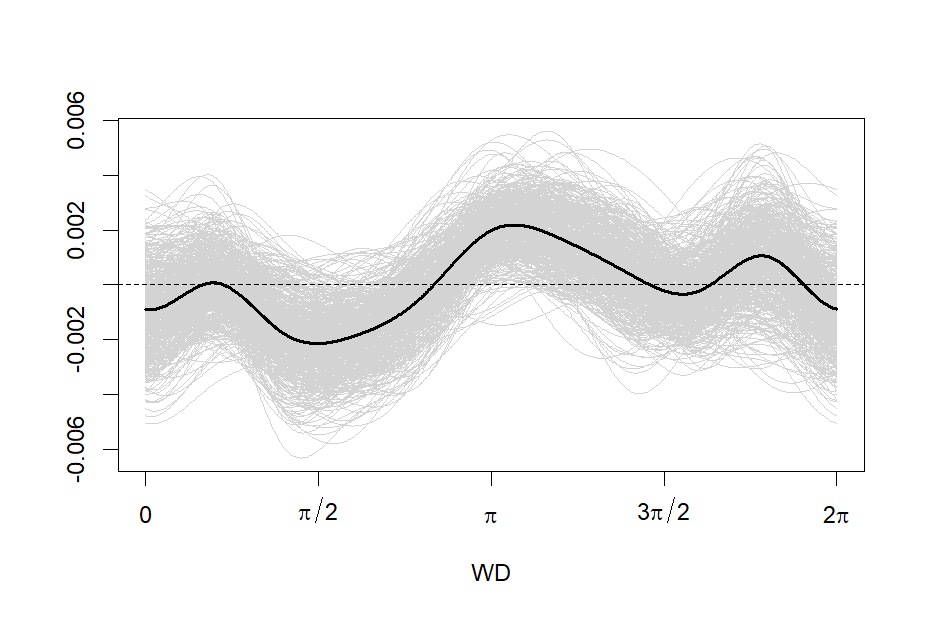}
      \caption{ Bootstrap confidence bands for functional time regression parameters in $L_0^2(0,2\pi)$ (intercept $\mathrm{clr} (\hat{ \beta_0}(t))$ (left) and time parameter $\mathrm{clr} (\hat{ \beta_1}(t))$ (right))}
      \label{fig:regT_boot} 
\end{figure}

The uncertainty analysis for the estimation of the functional regression parameters is performed by constructing confidence bands using the bootstrap method based on resampling the residuals of the model (see more details in \cite{talska18,talska21}). A total of 500 bootstrap estimates of the time regression parameters form the gray confidence bands illustrated in Figure~\ref{fig:regT_boot}. The left-hand panel shows low uncertainty in the estimation of the time intercept $\mathrm{clr} (\hat{ \beta_0}(t))$, as the confidence band follows the predicted curve closely throughout the entire domain. On the right-hand side, the confidence band for the time parameter $\mathrm{clr} (\hat{ \beta_1}(t))$ is considerably wider. This leads to the conclusion that the annual distribution of wind direction does not depend linearly on time and does not change significantly over months and years.

\begin{figure}[htbp]
    \centering
       \includegraphics[width=0.45\textwidth]{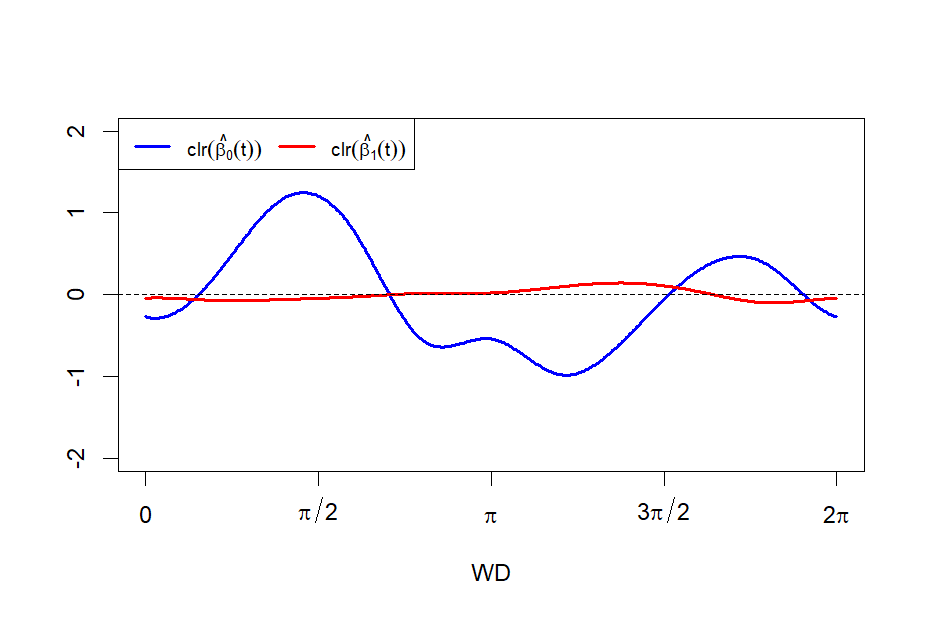}
    \hfill
    \includegraphics[width=0.45\textwidth]{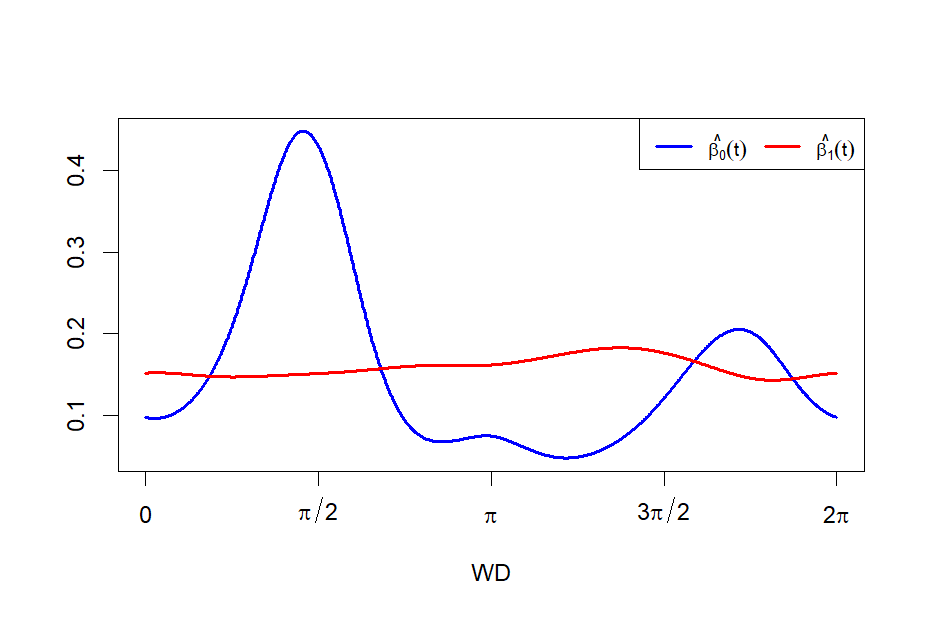} \\
    \includegraphics[width=0.45\textwidth]{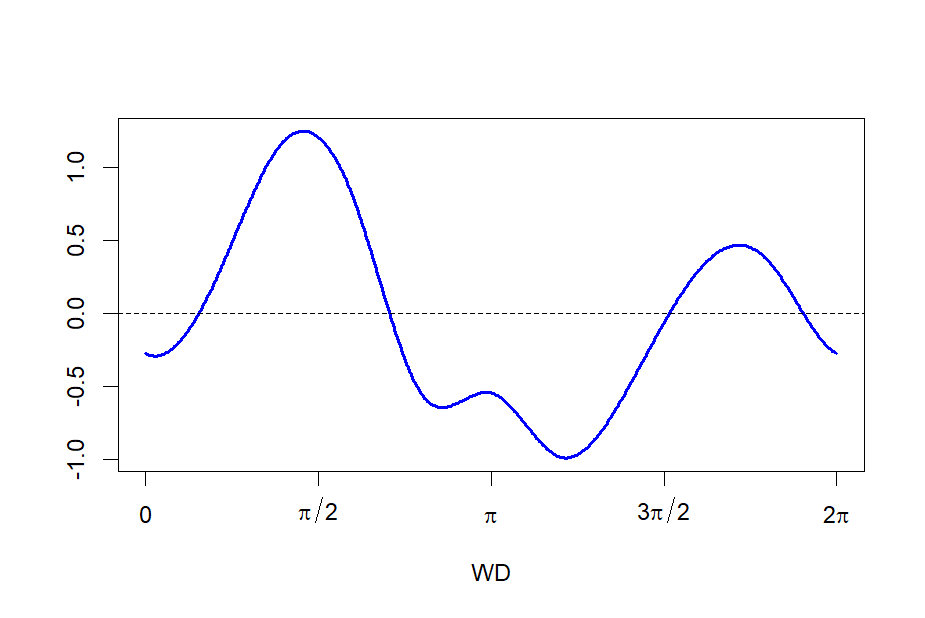}
    \hfill
    \includegraphics[width=0.45\textwidth]{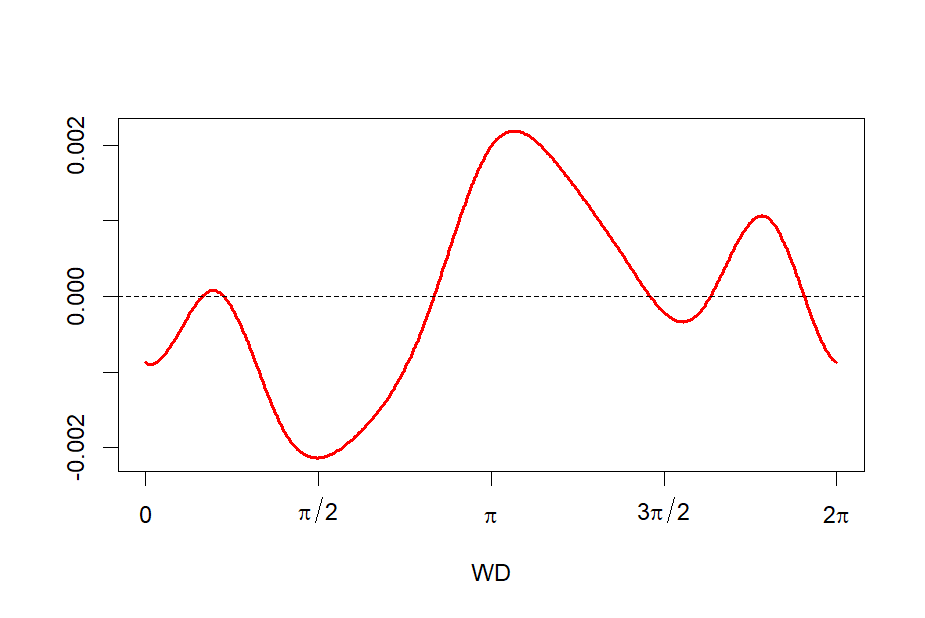}
      \caption{Functional wind speed regression parameters in $L_0^2(0,2\pi)$ resp. $\mathcal{B}^2(0,2\pi)$ (upper left resp. right); intercept $\mathrm{clr} (\hat{ \beta_0}(t))$ in $L_0^2(0,2\pi)$ (bottom left) and wind speed parameter $\mathrm{clr} (\hat{ \beta_1}(t))$ in $L_0^2(0,2\pi)$ (bottom right)}
      \label{fig:regWS_coef} 
\end{figure}

Last but not least, we examined the dependence of wind direction on wind speed, again using function-on-scalar regression. The explanatory variable values $ x_i$ and $i=1,\dots,120$ are the average monthly wind speeds in km/hr, calculated from the original dataset. 
The corresponding function-on-scalar regression model
in $\mathcal{B}^2(0,2\pi)$ can be written as
\begin{equation}\label{regWS_modelB}
y_i (t)  \, = \, \beta_0(t) \, \oplus \, [x_i \odot \beta_1](t) \, \oplus \, \varepsilon_i (t) \quad  i=1, \dotsc,120 , \quad t \in (0,2\pi),
\end{equation}
where $y_i (t)$ are wind direction densities, $\beta_0(t)$ and $ \beta_1(t)$ are wind speed regression parameters, and $\varepsilon_i (t)$ are random functional errors in $\mathcal{B}^2(0,2\pi)$. In $L_0^2(0,2\pi)$, all clr-transformed functions can be represented by periodic spline coefficients, and model~(\ref{regWS_modelB}) can be written as a multivariate model for the B-spline coefficients \citep{talska18}. 

\begin{figure}[htbp]
    \centering
    \includegraphics[width=0.45\textwidth]{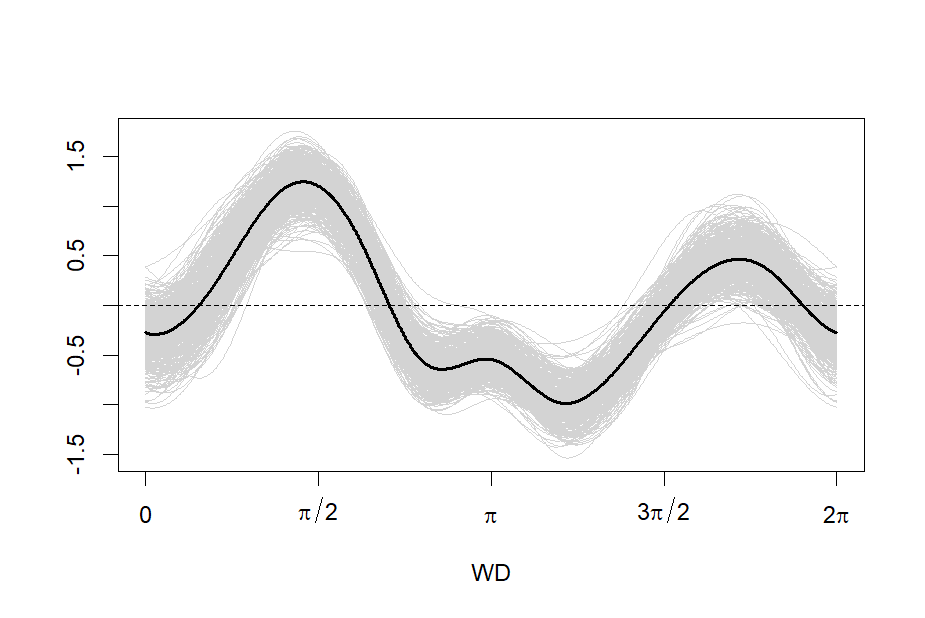}
    \hfill
    \includegraphics[width=0.45\textwidth]{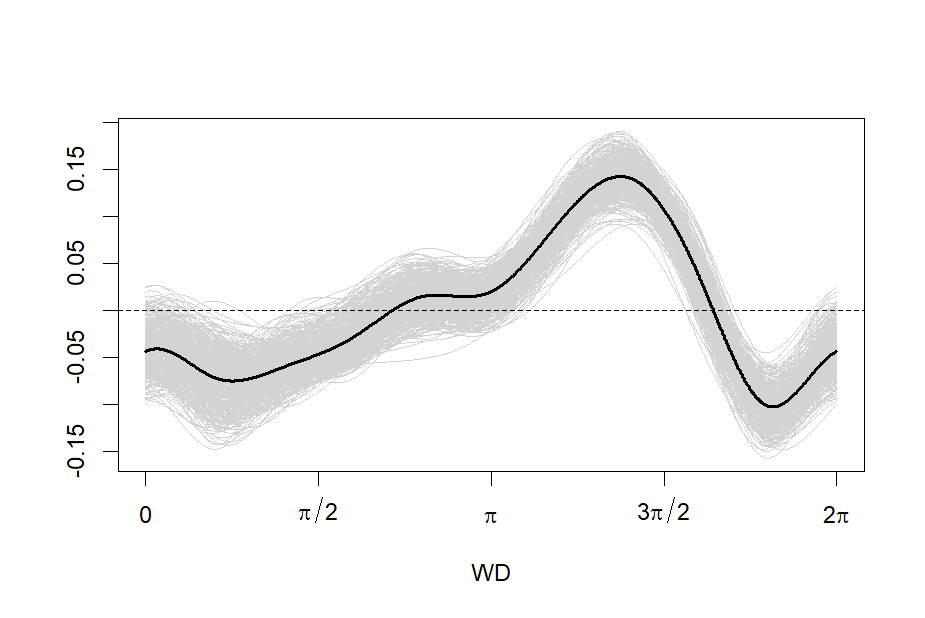}
      \caption{Bootstrap confidence bands for functional wind speed regression parameters in $L_0^2(0,2\pi)$ (intercept $\mathrm{clr} (\hat{ \beta_0}(t))$ (left) and wind speed parameter $\mathrm{clr} (\hat{ \beta_1}(t))$ (right))}
      \label{fig:regWS_boot} 
\end{figure}

\begin{figure}[htbp]
    \centering
    \includegraphics[width=0.45\textwidth]{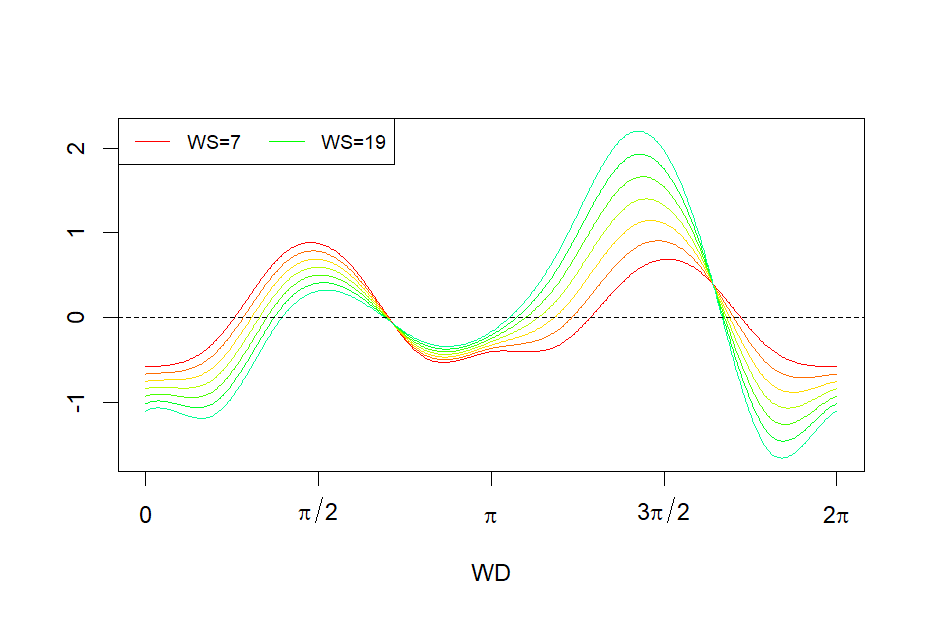}
    \hfill
    \includegraphics[width=0.45\textwidth]{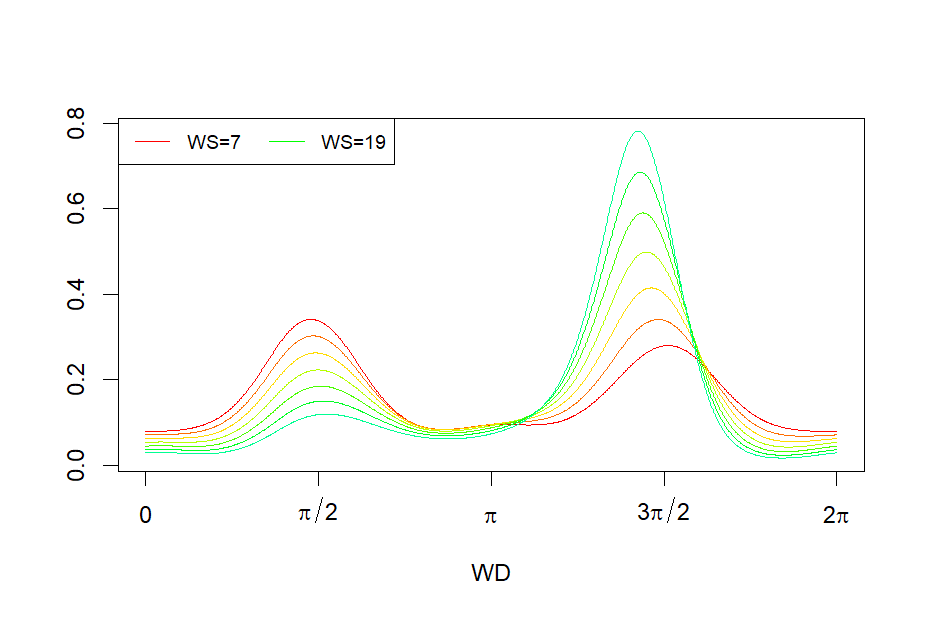}
      \caption{ Prediction of direction for wind speed ranging from 7 to 19 km/hr in $L_0^2(0,2\pi)$ (left) and corresponding wind direction densities in  $\mathcal{B}^2(0,2\pi)$ (right).}
      \label{fig:regWS_grow} 
\end{figure}

Similarly to the time regression, Figure~\ref{fig:regWS_coef} shows both functional wind speed regression parameters in $L_0^2(0,2\pi)$ in the upper left panel and the corresponding densities in $\mathcal{B}^2(0,2\pi)$ after inverse clr transformation in the upper right panel. The lower panels visualize the individual shapes of the intercept $\mathrm{clr} (\hat{ \beta_0}(t))$ and the wind speed parameter $\mathrm{clr} (\hat{ \beta_1}(t))$ in $L_0^2(0,2\pi)$. The uncertainty analysis for the estimation of the functional regression parameters is again performed using bootstrap confidence bands based on resampling the residuals of the model, as illustrated in Figure~\ref{fig:regWS_boot}. The influence of wind speed $\beta_1(t)$ is statistically significant, as is the intercept $\beta_0(t)$, and therefore it is meaningful to use the regression model to predict wind direction densities. Southwesterly wind is positively influenced by wind strength, while the north-to-east wind direction occurs less frequently as wind strength increases. This is confirmed by Figure~\ref{fig:regWS_grow}, where the left panel shows predictions for wind strength ranging from a minimum average of 7 km/hr (red) to a maximum of 19 km/hr (green) in the $L_0^2(0,2\pi)$ space, and the right panel shows the corresponding wind direction densities in $\mathcal{B}^2(0,2\pi)$ after inverse clr transformation. The numerical results confirm the ordinary climatic situation above Pardubice Airport, where strong winds tend to blow from the east and a gentle breeze typically comes from the west.

\section{Conclusion}
\label{sec:6}

This paper developed a compositional periodic spline framework for the approximation and analysis of circular density data in Bayes spaces. The proposed methodology combines the geometric structure of $\mathcal{B}^2(I)$ with periodic spline smoothing in the clr-transformed space $L_0^2(I)$, thereby respecting both the relative nature of density functions and the periodicity inherent to circular variables. In this way, the approach provides a statistically coherent and computationally feasible representation of directional distributions.

From the methodological point of view, the main contribution lies in the construction of periodic splines with zero integral, together with their matrix representation and the derivation of corresponding smoothing spline and P-spline estimators. This formulation allows one to work directly in a spline coefficient framework while preserving the structural constraints induced by the clr transformation. As a result, the proposed approach is suitable not only for smooth approximation of circular densities but also for subsequent statistical processing within the standard toolbox of functional data analysis. On the other hand, in contrast to the non-periodic setting, where $Z\!B$-spline bases provide a convenient representation of zero-integral functions \citep{machalova21}, such a construction is not compatible with periodic boundary conditions and therefore cannot be applied on circular domains.

The practical performance of the method was demonstrated on a long-term wind direction dataset from Pardubice Airport. The results showed that periodic spline approximation provides smooth and interpretable representations of monthly wind direction distributions, while the comparison of smoothing spline and P-spline variants indicated only minor differences in fit, with the first-order smoothing spline yielding the best overall performance according to the SSE criterion. The subsequent functional data analysis revealed a stable annual structure in the wind direction distributions, with no substantial linear temporal trend over the observed period. In contrast, the regression analysis with wind speed as an explanatory variable indicated a statistically meaningful relationship, confirming that stronger winds are associated with systematic changes in the directional distribution.

The proposed framework is not limited to wind direction data. More generally, it offers a basis for the analysis of periodic density-valued observations arising in a wide range of scientific applications. An important direction for future research is the extension to bivariate and multivariate density settings involving circular as well as non-circular variables. In such cases, the Bayes space representation may provide a natural way to separate main effects and interaction structures and thus support more detailed decomposition of multivariate distributions \citep{genest23}. A further promising line of development is the extension of the present methodology to densities defined on the sphere or, more generally, on manifolds, where periodicity is replaced by more complex geometric constraints. These perspectives suggest that the Bayes space approach has considerable potential for broader distributional modeling beyond the univariate circular case.

\subsection*{\bf{Declaration of competing interest}}

The authors declare that they have no known competing financial interests or personal relationships that could have
appeared to influence the work reported in this paper.

\subsection*{\bf{Funding}}
The work was supported by the Czech Science Foundation within the
project 25-15447S and by the institutional support of the University of Pardubice, Czech Republic.

%\paragraph*{\bf{Declaration of generative AI and AI-assisted technologies in the manuscript preparation process}}
%During the preparation of this work the authors used Copilot in order to literature review. After using this tool/service, the authors reviewed and edited the content as needed and take full responsibility for the content of the published article.

\bibliographystyle{chicago}
\bibliography{ref}

\end{document}